\documentclass[twocolappendix,11pt,a4paper]{emulateapj}

\usepackage{psfrag}
\usepackage{psfig}
\usepackage{pspicture}
\usepackage{graphicx}

\usepackage{amssymb,amsmath}
\usepackage{natbib}

\def\apj{ApJ}

\submitted{published in the \apj}

\begin{document}

\title {Similar Scaling Relations for the Gas Content of Galaxies across Environments to \lowercase{$z\sim$} 3.5}

\author{
Behnam Darvish\altaffilmark{1}, 
Nick Z. Scoville\altaffilmark{1},
Christopher Martin\altaffilmark{1},
Bahram Mobasher\altaffilmark{2},
Tanio Diaz-Santos\altaffilmark{3},
and Lu Shen\altaffilmark{4}
}

\setcounter{footnote}{0}
\altaffiltext{1}{Cahill Center for Astrophysics, California Institute of Technology, 1216 East California Boulevard, Pasadena, CA 91125, USA; email: bdarv@caltech.edu}
\altaffiltext{2}{University of California, Riverside, 900 University Avenue, Riverside, CA 92521, USA}
\altaffiltext{3}{Ncleo de Astronoma de la Facultad de Ingeniera, Universidad Diego Portales, Avenida Ejrcito Libertador 441, Santiago, Chile}
\altaffiltext{4}{Physics Department, University of California, Davis, One Shields Avenue, Davis, CA 95616, USA}
\begin{abstract}

We study the effects of the local environment on the molecular gas content of a large sample of log($M_{*}$/$M_{\odot}$) $\gtrsim$ 10 star-forming and starburst galaxies with specific star-formation rates (sSFRs) on and above the main sequence (MS) to $z$ $\sim$ 3.5. ALMA observations of the dust continuum in the COSMOS field are used to estimate molecular gas masses at $z$ $\approx$ 0.5-3.5. We also use a local universe sample from the ALFALFA HI survey after converting it into molecular masses. The molecular mass ($M_{ISM}$) scaling relation shows a dependence on $z$, $M_{*}$, and sSFR relative to the MS, but no dependence on environmental overdensity $\Delta$ ($M_{ISM}$ $\propto$ $\Delta^{0.03}$). Similarly, gas mass fraction (f$_{gas}$) and depletion timescale ($\tau$) show no environmental dependence to $z$ $\sim$ 3.5. At $\langle z\rangle$ $\sim$ 1.8, the average $\langle M_{ISM}\rangle$,$\langle$f$_{gas}\rangle$, and $\langle \tau \rangle$ in densest regions is (1.6$\pm$0.2)$\times$10$^{11}$ $M_{\odot}$, 55$\pm$2\%, and 0.8$\pm$0.1 Gyr, respectively, similar to those in the lowest density bin. Independent of the environment, f$_{gas}$ decreases and $\tau$ increases with increasing cosmic time. Cosmic molecular mass density ($\rho$) in the lowest density bins peaks at $z$ $\sim$ 1-2, and this peak happens at $z$ $<$ 1 in densest bins. This differential evolution of $\rho$ across environments is likely due to the growth of the large-scale structure with cosmic time. Our results suggest that the molecular gas content and the subsequent star-formation activity of log($M_{*}$/$M_{\odot}$) $\gtrsim$ 10 star-forming and starburst galaxies is primarily driven by internal processes, and not by their local environment since $z$ $\sim$ 3.5.
    
\end{abstract}


\section{Introduction} \label{int}

It is now well established that the environment of galaxies influences their properties such as morphology, color, star-formation rate (SFR), and likely metallicity, stellar mass, dust, and gas content (e.g., \citealp{Dressler80,Kauffmann04,Peng10,Scoville13,Darvish16,Sobral16}). 
  
Low-$z$ studies find a deficiency of atomic hydrogen in denser regions compared to the field \citep{Cayatte90,Gavazzi05,Cortese11,Serra12,Catinella13,Jaffe15,Brown17}. These studies along with observations of, e.g., jelly-fish galaxies in clusters \citep{Owers12,Poggianti16} suggest that this is likely due to the ram-pressure stripping the interstellar medium of galaxies. 

The situation is different for the molecular gas as it is less extended and much denser than HI and hence more bound to the host galaxy. A number of studies support this picture, finding no environmental effects on the molecular gas in galaxies \citep{Stark86,Kenney89,Lavezzi98,Koyama17}. However, others find a deficit \citep{Fumagalli09,Corbelli12,Jablonka13,Scott13,Boselli14c} or even an enhancement \citep{Mok16} of molecular gas in denser environments than the field.

Thanks to modern radio interferometers, the molecular gas studies based on CO observations are now performed to $z$ $\sim$ 3-4 (e.g., \citealp{Daddi10,Saintonge11a,Tacconi18}), suggesting a higher gas mass fraction at higher $z$ and other potential dependence of the molecular gas on stellar mass (lower gas fraction in more massive systems) and SFR relative to the main sequence (higher gas fraction for systems above the main sequence; e.g., \citealp{Tacconi13,Genzel15}). However, these studies often target field galaxies, and statistically large samples of high-$z$ galaxies in dense environments are essential.

\begin{figure*}
\centering
\includegraphics[width=7.0in]{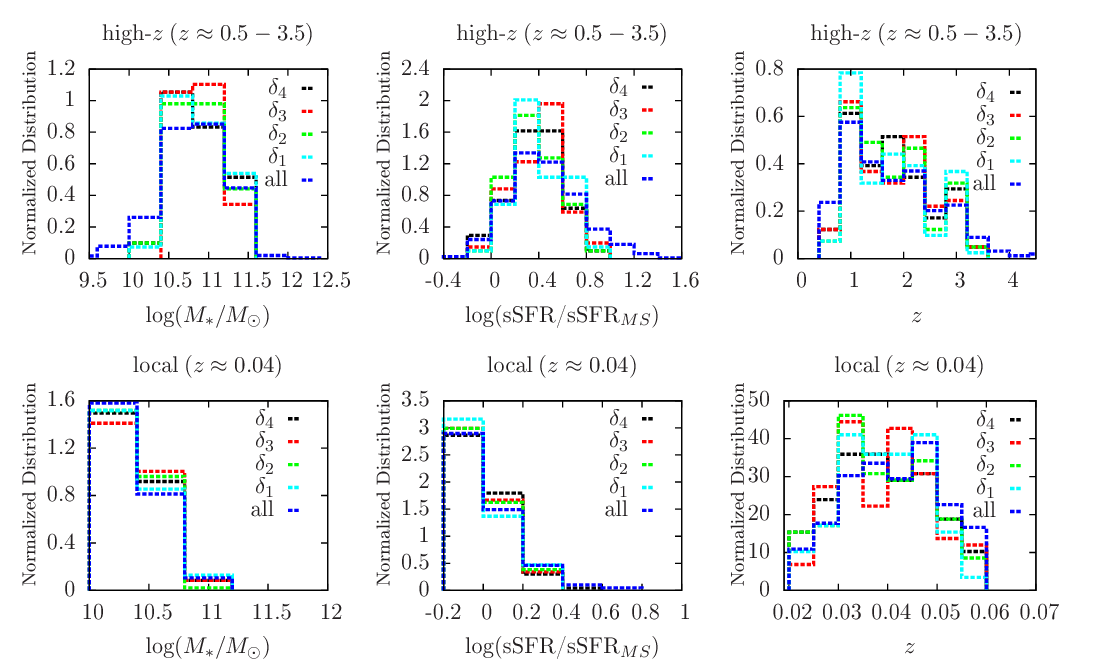}
\caption{Top: Distribution of stellar mass, sSFR relative to the main sequence, and redshift in four environmental density quantiles of $\delta_{4}$ (densest), $\delta_{3}$, $\delta_{2}$, and $\delta_{1}$ for our high-$z$ sample ($z$ $\approx$ 0.5-3.5). The parent high-$z$ sample from \cite{Scoville17} is also shown with blue dashed lines. Note the similarity of distributions in different density quantiles. In all cases, K-S test p-value is $>$ 0.1, ensuring a good control sample selection (Section \ref{control}). High-$z$ samples are located on and above the main sequence and have $M_{*}$ $\gtrsim$ 10$^{10}$ $M_{\odot}$. Bottom: Similar to the top panels, but for our local universe sample at $z$ $\sim$ 0.04. Note the similarity of $M_{*}$, sSFR/sSFR$_{MS}$, and $z$ distributions in different environmental density bins, showing a good control sample selection (Section \ref{control}). Similar to our high-$z$ sample, our local universe galaxies are located on and above the main sequence with $M_{*}$ $\geqslant$ 10$^{10}$ $M_{\odot}$.}
\label{fig:hist-sct-all}
\end{figure*}

Nonetheless, there are a limited number of molecular gas studies in dense environments at high $z$, often with limited CO detections \citep{Aravena12,Wagg12,Hayashi17,Lee17,Noble17,Rudnick17,Hayashi18} and often with contradictory results, likely because of small sample size, different selections, and a small dynamical range in stellar mass ($M_{*}$) and the sSFR with respect to the main sequence (sSFR/sSFR$_{MS}$). For example, \cite{Noble17} find a higher gas fraction and a similar depletion timescale in $z$ $\sim$ 1.6 clusters than the field, \cite{Hayashi18} find a higher gas fraction and depletion timescale for a $z$=1.46 cluster galaxies than the field, and \cite{Lee17} find a similar gas fraction but likely shorter depletion timescale in denser regions at $z$ $\sim$ 2.5. 

One major problem with molecular mass estimates based on CO observations is that they are time-consuming, often leading to a small sample size. Moreover, converting the excited state CO fluxes into reliable estimates of the gas content in galaxies remains an issue \citep{Carilli13}. 

A much faster (minute-long versus hour-long) alternative arises from the long-wavelength Rayleigh-Jeans (RJ) dust-continuum method (e.g., \citealp{Santini14,Scoville14}). This technique has been calibrated over a broad range of CO luminosities \citep{Scoville14}, successfully leading to ISM mass estimates for 708 galaxies at $z$ $\sim$ 0.3-4.5 with ALMA \citep{Scoville17}. This large blind sample increases the chance of finding galaxies that happen to be in dense regions, even though the original selection is not designed to target dense environments \textit{a priori}. 

\cite{Scoville17} studied the molecular gas content of galaxies as a function of $z$, $M_{*}$, and sSFR/sSFR$_{MS}$, providing analytical expressions for gas mass, gas fraction, depletion timescale, etc. The missing parameter is the environment, which will be investigated here. 

In this paper, by using large samples of galaxies covering a broad range of environments, $z$, $M_{*}$, and sSFR/sSFR$_{MS}$, we address the following question: Given fixed $z$, $M_{*}$, and sSFR/sSFR$_{MS}$, does the local environment affect the molecular gas content of galaxies in the local universe and at higher redshift? For the high-$z$ sample, we rely on the \cite{Scoville17} large sample by dividing it into environmental bins. For the local universe, we rely on the ALFALFA HI sample \citep{Haynes11} after converting it into molecular gas and placing the galaxies into different density bins.                 

In Section \ref{data} we present the data. The methods used to quantify the local environment of galaxies are developed in Section \ref{env}. The control sample selection is given in Section \ref{control}. Our results are presented in Section \ref{result}, discussed in Section \ref{dis}, and summarized in Section \ref{sum}. Throughout this work, we assume a flat $\Lambda$CDM cosmology with $H_{0}$=70 kms$^{-1}$ Mpc$^{-1}$, $\Omega_{m}$=0.3, and $\Omega_{\Lambda}$=0.7, a Chabrier initial mass function (IMF; \citealp{Chabrier03}), and a CO-to-H$_{2}$ conversion factor of $\alpha_{CO}$=6.5 $M_{\odot}$ (K kms$^{-1}$pc$^{2}$)$^{-1}$ \citep{Scoville14}. The relevant values from the literature are converted into our adopted IMF and $\alpha_{CO}$.

\begin{table*}
\begin{center}
\caption{Density ($\Sigma$) and Overdensity ($\Delta$) Values for Different Environmental Density Quantiles} 
\begin{scriptsize}
\centering
\begin{tabular}{lccccccccc}
\hline
\noalign{\smallskip}
Redshift & $\Sigma(\delta_{1})$\footnote{Density range for the $\delta_{1}$ quantile.} & $\Sigma(\delta_{2})$ & $\Sigma(\delta_{3})$ & $\Sigma(\delta_{4})$ & $\Delta$($\delta_{1}$)\footnote{Overdensity range for the the $\delta_{1}$ quantile. Overdensity is defined as the density divided by the median density at each redshift.} & $\Delta$($\delta_{2}$) & $\Delta$($\delta_{3}$) & $\Delta$($\delta_{4}$)\\
& (Mpc$^{-2}$) & (Mpc$^{-2}$) & (Mpc$^{-2}$) & (Mpc$^{-2}$) & & & &\\
\hline
\\
0.02$-$0.12 & $\leqslant$ 0.47 & 0.47$-$1.03 & 1.03$-$2.76 & $\geqslant$ 2.76 & $\leqslant$ 0.45 & 0.45$-$1.00 & 1.00$-$2.69 & $\geqslant$ 2.69\\
\\
0.2$-$0.5 & $\leqslant$ 1.01 & 1.01$-$2.26 & 2.26$-$4.42 & $\geqslant$ 4.42 & $\leqslant$ 0.46 & 0.46$-$1.00 & 1.00$-$2.02 & $\geqslant$ 2.02\\
\\
0.5$-$0.8 & $\leqslant$ 0.79 & 0.79$-$1.92 & 1.92$-$3.92 & $\geqslant$ 3.92 & $\leqslant$ 0.40 & 0.40$-$1.00 & 1.00$-$1.99 & $\geqslant$ 1.99\\
\\
0.8$-$1.1 & $\leqslant$ 0.85 & 0.85$-$1.89 & 1.89$-$3.55 & $\geqslant$ 3.55 & $\leqslant$ 0.45 & 0.45$-$1.00 & 1.00$-$1.89 & $\geqslant$ 1.89\\
\\
1.1$-$1.5 & $\leqslant$ 0.87 & 0.87$-$1.66 & 1.66$-$2.94 & $\geqslant$ 2.94 & $\leqslant$ 0.52 & 0.52$-$1.00 & 1.00$-$1.75 & $\geqslant$ 1.75\\
\\
1.5$-$2.0 & $\leqslant$ 0.93 & 0.93$-$1.62 & 1.62$-$2.65 & $\geqslant$ 2.65 & $\leqslant$ 0.57 & 0.57$-$1.00 & 1.00$-$1.62 & $\geqslant$ 1.62\\
\\
2.0$-$2.5 & $\leqslant$ 0.81 & 0.81$-$1.36 & 1.36$-$2.20 & $\geqslant$ 2.20 & $\leqslant$ 0.58 & 0.58$-$1.00 & 1.00$-$1.59 & $\geqslant$ 1.59\\
\\
2.5$-$3.0 & $\leqslant$ 0.56 & 0.56$-$1.10 & 1.10$-$1.96 & $\geqslant$ 1.96 & $\leqslant$ 0.50 & 0.50$-$1.00 & 1.00$-$1.71 & $\geqslant$ 1.71\\
\\
3.0$-$4.0 & $\leqslant$ 0.35 & 0.35$-$0.81 & 0.81$-$1.73 & $\geqslant$ 1.73 & $\leqslant$ 0.41 & 0.41$-$1.00 & 1.00$-$2.01 & $\geqslant$ 2.01\\
\\
\hline
\label{table1}
\end{tabular}
\end{scriptsize}
\end{center}
\end{table*}

\section{Data \& Sample Selection} \label{data}

\subsection{High Redshift} \label{data-HZ}

The parent sample is from \cite{Scoville17}, based on ALMA observations of the long-wavelength dust continuum for a sample of 708 galaxies at $z$ $\approx$ 0.3-4.5 in the COSMOS field \citep{Scoville07}. The sample galaxies have stellar masses of log($M_{*}$/$M_{\odot}$) $\gtrsim$ 10 and sSFRs from the main sequence to $\sim$ 50 times the MS (Figure \ref{fig:hist-sct-all}). Stellar masses are based on spectral energy distribution (SED) template fitting using optical to infrared bands \citep{Laigle16}. SFRs are estimated using the UV and far-infrared luminosity \citep{Lee13,Lee15a}. For the $M_{*}$ dependence of the MS, we use the shape from \cite{Lee15a} at $z$=1.2. For the redshift evolution of the MS, we use model \# 49 of \cite{Speagle14} evaluated at log($M_{*}$/$M_{\odot}$)=10.5. Molecular mass ($M_{ISM}$) estimates (corrected for heavy elements contribution) rely on the RJ dust-continuum emission in ALMA bands 6 and 7 (see \citealp{Scoville14,Scoville17} for details). Note that the parent sample selection is not designed \textit{a priori} to target galaxies in dense environments. Therefore, we estimate the local environment of the parent sample (Section \ref{env-HZ}) and then place the galaxies into four local density quantiles: $\delta_{1}$, $\delta_{2}$, $\delta_{3}$, and $\delta_{4}$ (highest density). This results in 102 galaxies in each density bin after controlling for their $M_{*}$, $z$, and sSFR/sSFR$_{MS}$ (Section \ref{control}). They cover the redshift range $z$ $\approx$ 0.5-3.5, log(sSFR/sSFR$_{MS}$) $\gtrsim$ -0.2, and log($M{*}$/$M_{\odot}$) $\gtrsim$ 10 (Figure \ref{fig:hist-sct-all}).   

\begin{figure*}
\centering
\includegraphics[width=7.0in]{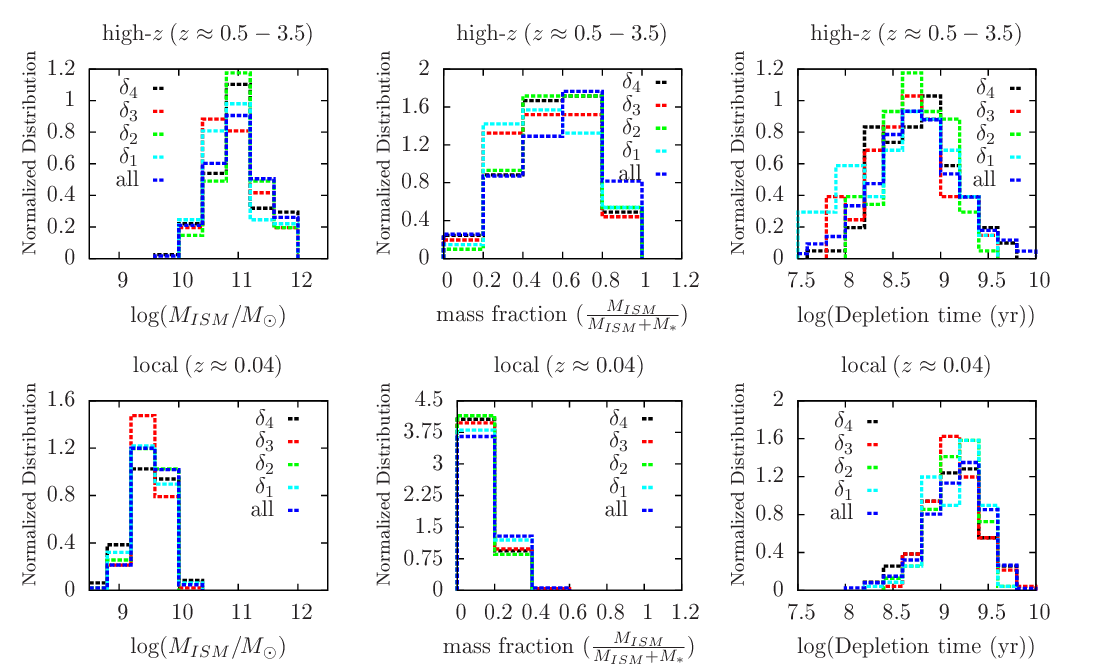}
\caption{Top: Distribution of molecular mass, gas mass fraction, and depletion timescale in different environmental density quantiles for our high-$z$ sample ($z$ $\approx$ 0.5-3.5), after controlling for $M_{*}$, sSFR/sSFR$_{MS}$, and $z$ in different density quantiles (Figure \ref{fig:hist-sct-all}). The parent high-$z$ sample from \cite{Scoville17} is also shown with blue dashed lines. Note the similarities between different density quantiles. We find no clear environmental dependence of $M_{ISM}$, f$_{gas}$, and $\tau$ for our high-$z$ sample. In all cases, the K-S test p-value is $>$ 0.1, showing that it is unlikely that the distributions of $M_{ISM}$, f$_{gas}$, and $\tau$ in different environmental density quantiles are drawn from different parent distributions. Bottom: Similar to the top panels, but for our local universe sample at $z$ $\sim$ 0.04. We do not find a significant environmental dependence of $M_{ISM}$, f$_{gas}$, and $\tau$ for our local universe sample, in line with our high-$z$ results.}
\label{fig:hist-all}
\end{figure*}

\subsection{Local Universe} \label{data-LO}

The local universe sample is from the ALFALFA HI survey \citep{Haynes11} matched with the SDSS DR12 \citep{Alam15}. Our SDSS galaxies have magnitudes $r$ $\leqslant$ 17.7, are in the spectroscopic redshift range of 0.02 $\leqslant$ $z$ $\leqslant$ 0.12, and they are located in the contiguous northern galactic cap (130.0 $\leqslant$ RA (deg) $\leqslant$ 240.0 and 0.0 $\leqslant$ Dec (deg) $\leqslant$ 60.0) \citep{Darvish18}. The mass completeness limit of the SDSS sample is log($M_{*}^{limit}/M_{\odot}$) $\sim$ 10.0 to $z$=0.12. To make our local universe selection as similar as possible to that of our high-$z$ sample, we further select galaxies that are on and above the MS (log(sSFR/sSFR$_{MS}$) $>$ -0.2) and with log($M{*}$/$M_{\odot}$) $\geqslant$ 10 (Figure \ref{fig:hist-sct-all}). The local universe MS is based on \cite{Salim07} (log(SFR)=0.65 log($M{*}$/$M_{\odot}$)$-$6.33). Stellar masses and SFRs are from the MPA-JHU DR8 catalog \citep{Kauffmann03}. We convert HI into H$_{2}$ mass assuming $M_{H_{2}}$/$M_{HI}$=0.26 \citep{Obreschkow09} and further multiply it by 1.36 to account for heavy element contribution \footnote{We note that the $M_{H_{2}}$/$M_{HI}$ ratio also depends on other parameters such as morphology, total gas, and $M_{*}$ (e.g., \citealp{Young91}). We also try $M_{H_{2}}$/$M_{HI}$ as a function of total gas mass (equation 12 in \cite{Obreschkow09}). Although the extracted gas-related parameters change, the overall results regarding the similarity of gas content in different environments (Section \ref{result}) remain the same.}. We assign our ALFALFA/SDSS matched galaxies to different local density quantiles (Section \ref{env-LO}), leading to 117 galaxies in each density quantile after controlling for their $M_{*}$, $z$, and sSFR/sSFR$_{MS}$ (Section \ref{control}). Figure \ref{fig:hist-sct-all} shows their redshift range ($z$ $\approx$ 0.02-0.06), sSFRs (log(sSFR/sSFR$_{MS}$) $>$ -0.2), and $M_{*}$ distribution (log($M{*}$/$M_{\odot}$) $\gtrsim$ 10).

\section{Environment Measurements} \label{env}

\subsection{High Redshift} \label{env-HZ}

We use the slightly modified density field estimation of \cite{Darvish17} in the COSMOS field. The local environment measurement relies on the adaptive kernel smoothing method \citep{Scoville13,Darvish15a} estimated over a series of overlapping redshift slices \citep{Darvish15a}. A mass-complete sample (log($M_{*}$/$M_{\odot}$) $\geqslant$ 10.4 similar to a volume-limited sample) is used for the density estimation. The lower limit of log($M_{*}$/$M_{\odot}$) $\geqslant$ 10.4 roughly corresponds to the stellar mass completeness limit at the highest redshift of this study \citep{Laigle16}. This selection is used to minimize the unrealistic underestimation of the density values at higher $z$ due to the Malmquist bias. The density values for our lower mass sample galaxies ( log($M_{*}$/$M_{\odot}$)=10-10.4) are estimated by interpolation of the density field to their positions. We define four density bins using the projected surface density quantiles from the low- to high-density regions: $\delta_{1}$, $\delta_{2}$, $\delta_{3}$, and $\delta_{4}$. The dynamical range of environments for our sources is $\sim$ 2.1 dex, sampling those of the overall galaxy population. The range of density ($\Sigma$) and overdensity ($\Delta$) values for the defined density quantiles at different redshifts are given in Table \ref{table1}. Overdensity ($\Delta$) is defined as the density divided by the median density at each redshift (see, e.g., \citealp{Darvish17}).

To compare the overdensity values with those of real galaxy groups/clusters, we use a sample of X-ray groups at $z$ $<$ 1.2 in the COSMOS field with measured positions, $z$, $R_{200}$, and $M_{200}$ from \cite{Finoguenov07}. For each reliable group (flag=1), we determine the median overdensity of galaxies (used for density estimation) within $R_{200}$ of each group center and in redshift slices around the $z$ of each group. The width of each redshift slice is similar to those used here for density estimation (also see \citealp{Darvish17}). For a median $M_{200}$=4.0 $\times$ 10$^{13}$ $M_{\odot}$ and 2.1 $\times$ 10$^{14}$ $M_{\odot}$ group, we obtain a median overdensity of 3.4 and 10.2, respectively. Unfortunately, no such catalog exists at $z$ $>$ 1.2 for comparison with our estimated density values.
      
\subsection{Local Universe} \label{env-LO}

We use the density measurements of \cite{Darvish18} based on the SDSS DR12 with redshift, magnitude, and angular position cuts already defined in Section \ref{data-LO}. Local density estimations are based on the projected comoving distance to the $10th$ nearest neighbor to each galaxy, considering only galaxies that are within the recessional velocity range of $\Delta v$=$c\Delta z$=$\pm$1000 kms$^{-1}$ to the galaxy of interest, and corrected for incompleteness due to the fiber collision and the Malmquist bias. Similar to the high-$z$ sample, we place galaxies into four density bins. The dynamical range of environments here is $\sim$ 2.9 dex. Table 1 lists the range of density and overdensity values for the density quantiles.

Using different density estimators at high- and low-$z$ (adaptive kernel smoothing versus $10th$ nearest neighbor) might lead to a potential bias in presenting the low- and high-$z$ results. However, \cite{Darvish17} compared the density values with the $10th$ nearest neighbor and adaptive kernel smoothing methods for a similar sample at high-$z$ and found a good agreement (an offset of $<$ 0.1 dex and a median absolute deviation of $<$ 0.2 dex). Moreover, \cite{Darvish15a} find an overall good agreement between the estimated density fields using different methods (including the adaptive kernel smoothing and $10th$ nearest neighbor) over $\sim$ 2 dex in overdensity values through simulations and also observational data. Hence, the selection of different estimators does not likely have a significant effect on the presented results. 

\section{Control Sample Selection} \label{control} 

Since the estimated $M_{ISM}$ of galaxies likely depends on $z$, $M_{*}$, and sSFR/sSFR$_{MS}$ (e.g., \citealp{Genzel15,Scoville17}), we control for these parameters when we compare our high-density sample ($\delta_{4}$) with the rest. To do this, we use our high-density sample as the reference \footnote{In principle, any density quantile could be used as the reference. For a sanity check, we also try the lowest density sample and find similar results.}. Then, for each galaxy in this sample, we first search for any galaxy in the lower-density samples (e.g., $\delta_{3}$) that is within $\pm$ 0.2 dex of $M_{*}$, $\pm$ 0.2 dex of sSFR/sSFR$_{MS}$, and $\pm$ 0.2 ($\pm$0.005 for the local sample) of redshift of that galaxy. If more than one match is found, we then select the galaxy that has the closest Cartesian distance to the galaxy of our interest in the 3D $M_{*}$-sSFR/sSFR$_{MS}$-$z$ space. If no match is found, we remove that galaxy from the high-density sample. This results in 102 (117) galaxies in each density bin for our high-$z$ (local universe) sample. 

\begin{table*}
\begin{center}
\caption{Properties of Samples in Different Environmental Density Quantiles} 
\begin{scriptsize}
\centering
\begin{tabular}{lccccccccc}
\hline
\noalign{\smallskip}
Sample\footnote{$\delta_{1}$, $\delta_{2}$, $\delta_{3}$, and $\delta_{4}$ refer to density quantiles from the low- to high-density regions, respectively.} & N\footnote{Sample size} & $\langle z \rangle$\footnote{Average redshift} & $\langle M_{ISM}\rangle$\footnote{Average molecular mass, assuming $\alpha_{CO}$=6.5 $M_{\odot}$ (K kms$^{-1}$pc$^{2}$)$^{-1}$} & $M_{ISM,med}$\footnote{Median molecular mass} & $\langle$f$_{gas}\rangle$\footnote{Average gas mass fraction} & f$_{gas,med}$\footnote{Median gas mass fraction} & $\langle \tau \rangle$\footnote{Average depletion timescale} & $\tau_{med}$\footnote{Median depletion timescale}\\
& & & (10$^{11}$ $M_{\odot}$) & (10$^{11}$ $M_{\odot}$) & & & (Gyr) & (Gyr)\\
\hline
\\
local($\delta_{1}$)(all) & 	117 & 0.0395 & 0.039$\pm$0.002 & 0.035$\pm$0.002 & 0.145$\pm$0.006 & 0.136$\pm$0.007 & 1.3$\pm$0.1 & 1.2$\pm$0.1\\
\\
local($\delta_{2}$)(all) &	117 & 0.0391 & 0.039$\pm$0.002 & 0.034$\pm$0.002 & 0.141$\pm$0.005 & 0.140$\pm$0.005 & 1.2$\pm$0.1 & 1.0$\pm$0.1\\
\\
local($\delta_{3}$)(all) & 	117 & 0.0394 & 0.037$\pm$0.002 & 0.033$\pm$0.002 & 0.148$\pm$0.007 & 0.129$\pm$0.005 & 1.4$\pm$0.2 & 0.9$\pm$0.2\\
\\
local($\delta_{4}$)(all) & 	117 & 0.0390 & 0.039$\pm$0.002 & 0.035$\pm$0.002 & 0.139$\pm$0.007 & 0.130$\pm$0.006 & 1.3$\pm$0.1 & 1.0$\pm$0.1\\
\\
\hline
\\
high-$z$($\delta_{1}$)(all) &	102 & 1.77 & 1.4$\pm$0.2 & 0.8$\pm$0.2 & 0.52$\pm$0.02 & 0.52$\pm$0.02 & 0.7$\pm$0.1 & 0.5$\pm$0.1\\
\\
high-$z$($\delta_{2}$)(all) &	102 & 1.79 & 1.4$\pm$0.1 & 1.1$\pm$0.1 & 0.56$\pm$0.02 & 0.56$\pm$0.02 & 0.7$\pm$0.1 & 0.6$\pm$0.1\\
\\
high-$z$($\delta_{3}$)(all) & 	102 & 1.80 & 1.5$\pm$0.2 & 0.8$\pm$0.2 & 0.53$\pm$0.02 & 0.52$\pm$0.02 & 0.7$\pm$0.1 & 0.5$\pm$0.1\\
\\
high-$z$($\delta_{4}$)(all) &	102 & 1.79 & 1.6$\pm$0.2 & 1.0$\pm$0.2 & 0.55$\pm$0.02 & 0.57$\pm$0.02 & 0.8$\pm$0.1 & 0.6$\pm$0.1\\
\\
\hline
\\
high-$z$($\delta_{1}$)($z$ $<$ 1.5) & 	38 & 1.04 & 0.7$\pm$0.1 & 0.6$\pm$0.1 & 0.46$\pm$0.03 & 0.44$\pm$0.04 & 1.1$\pm$0.1 & 1.0$\pm$0.1\\
\\
high-$z$($\delta_{2}$)($z$ $<$ 1.5) & 	38 & 1.05 & 0.6$\pm$0.1 & 0.7$\pm$0.1 & 0.44$\pm$0.03 & 0.45$\pm$0.04 & 1.1$\pm$0.1 & 1.0$\pm$0.1\\
\\
high-$z$($\delta_{3}$)($z$ $<$ 1.5) & 	38 & 1.06 & 0.6$\pm$0.1 & 0.5$\pm$0.1 & 0.42$\pm$0.03 & 0.40$\pm$0.03 & 1.0$\pm$0.1 & 0.8$\pm$0.1\\
\\
high-$z$($\delta_{4}$)($z$ $<$ 1.5) & 	38 & 1.07 & 0.6$\pm$0.1 & 0.5$\pm$0.1 & 0.43$\pm$0.03 & 0.42$\pm$0.03 & 1.1$\pm$0.2 & 0.7$\pm$0.2\\
\\
\hline
\\
high-$z$($\delta_{1}$)(1.5 $\leqslant$ $z$ $<$ 2.5) &	37 & 1.94 & 1.8$\pm$0.4 & 1.1$\pm$0.4 & 0.54$\pm$0.03 & 0.57$\pm$0.03 & 0.6$\pm$0.1 & 0.4$\pm$0.1\\
\\
high-$z$($\delta_{2}$)(1.5 $\leqslant$ $z$ $<$ 2.5) & 	37 & 1.93 & 1.7$\pm$0.3 & 1.2$\pm$0.4 & 0.61$\pm$0.03 & 0.63$\pm$0.04 & 0.7$\pm$0.1 & 0.5$\pm$0.1\\
\\
high-$z$($\delta_{3}$)(1.5 $\leqslant$ $z$ $<$ 2.5) & 	37 & 2.00 & 2.0$\pm$0.4 & 1.0$\pm$0.4 & 0.58$\pm$0.03 & 0.55$\pm$0.03 & 0.6$\pm$0.1 & 0.4$\pm$0.1\\
\\
high-$z$($\delta_{4}$)(1.5 $\leqslant$ $z$ $<$ 2.5) & 	37 & 1.94 & 2.0$\pm$0.3 & 1.0$\pm$0.4 & 0.62$\pm$0.03 & 0.63$\pm$0.03 & 0.7$\pm$0.1 & 0.5$\pm$0.1\\
\\
\hline
\\
high-$z$($\delta_{1}$)($z$ $\geqslant$ 2.5) & 	19 & 2.95 & 1.9$\pm$0.2 & 1.5$\pm$0.3 & 0.61$\pm$0.05 & 0.58$\pm$0.08 & 0.3$\pm$0.1 & 0.2$\pm$0.1\\
\\
high-$z$($\delta_{2}$)($z$ $\geqslant$ 2.5) &	19 & 2.97 & 2.4$\pm$0.3 & 1.8$\pm$0.4 & 0.68$\pm$0.04 & 0.67$\pm$0.06 & 0.3$\pm$0.1 & 0.3$\pm$0.1\\
\\
high-$z$($\delta_{3}$)($z$ $\geqslant$ 2.5) & 	19 & 2.95 & 2.5$\pm$0.6 & 1.7$\pm$0.6 & 0.64$\pm$0.05 & 0.72$\pm$0.04 & 0.3$\pm$0.1 & 0.2$\pm$0.1\\
\\
high-$z$($\delta_{4}$)($z$ $\geqslant$ 2.5) &	19 & 2.94 & 2.9$\pm$0.6 & 2.0$\pm$0.5 & 0.65$\pm$0.04 & 0.66$\pm$0.05 & 0.4$\pm$0.1 & 0.3$\pm$0.1\\
\\
\hline
\label{table2}
\end{tabular}
\end{scriptsize}
\end{center}
\end{table*}

Figure \ref{fig:hist-sct-all} shows the normalized distributions of $M_{*}$, sSFR/sSFR$_{MS}$, and $z$ for the original high-density sample and all the other lower-density control samples for both the high-$z$ and local universe galaxies. Kolmogorov-Smirnov (K-S) tests are performed to ensure that distributions for different density bins are unlikely to be drawn from different parent distributions in $M_{*}$, sSFR/sSFR$_{MS}$, and $z$. High K-S p-values (p $>$ 0.1) in all cases guarantee a good control sample selection.
  
We note that this approach of controlling samples in different environments might lead to some bias and misinterpretation of the results if there is some degree of association between denser environments and stellar mass and/or the sSFR/sSFR$_{MS}$ of star-forming and starburst systems. Therefore, careful analysis is required prior to investigating potential differences between control samples in different environmental density bins. In the Appendix \ref{bias}, we discuss some relevant potential biases.                 

\begin{figure*}
\centering
\includegraphics[width=7.0in]{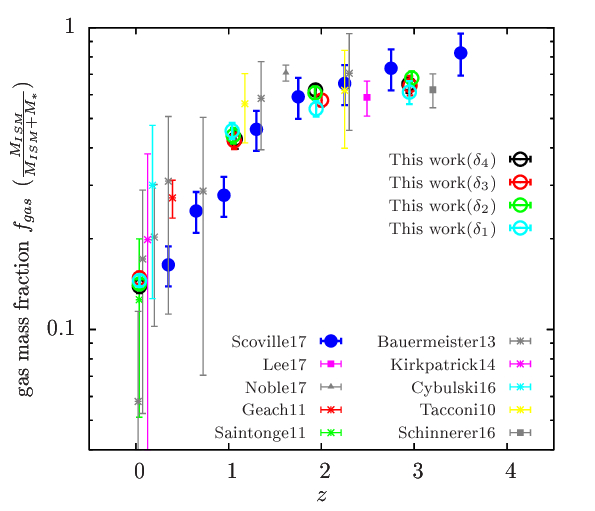}
\caption{Redshift evolution of gas mass fraction f$_{gas}$ in different environmental density quantiles (empty circles), along with cluster and field quantities in the literature. Note that the literature studies have their own sample selections that might be different than ours, and they are just shown for reference. f$_{gas}$ values in different environmental density bins show that independent of the environment, the gas mass fraction decreases with decreasing redshift and that our data points nearly follow the overall decline in f$_{gas}$ with cosmic time.}
\label{fig:fgas-z}
\end{figure*}

\begin{figure}
\centering
\includegraphics[width=3.5in]{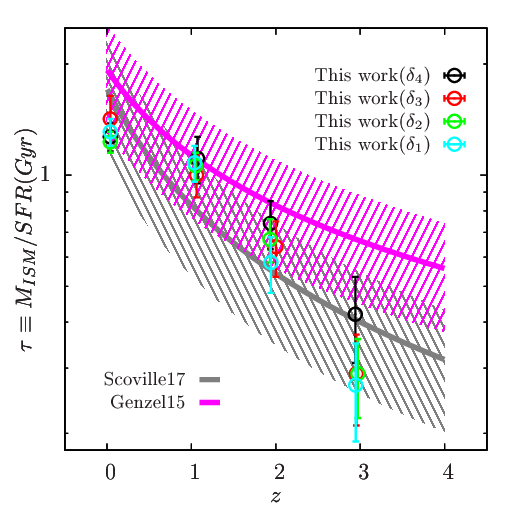}
\caption{Redshift evolution of depletion timescale $\tau$ in different environmental density quantiles (empty circles). Analytic expressions from \cite{Genzel15} and \cite{Scoville17} are shown. In these equations, we use the median sSFR/sSFR$_{MS}$ and $M_{*}$ of our sample. Independent of the environment, the depletion timescale increases with decreasing redshift. Within the uncertainties, our data points follow the overall increase in $\tau$ with cosmic time reasonably well.}
\label{fig:tau-z}
\end{figure}

\section{Results} \label{result}

\subsection{Scaling Relations with the Environment} \label{fits}

Prior to placing our sample galaxies into different density quantiles, we first fit a power-law dependence of $M_{ISM}$ of all galaxies as a function of $z$, sSFR/sSFR$_{MS}$, $M_{*}$ (similar to \citealp{Scoville17}), and overdensity ($\Delta$). For this analysis, we only rely on our ALMA dust-continuum high-$z$ sample, but similar results are expected for the local universe sample. To find the best-fit parameters, we use both a multidimensional nonlinear Levenberg-Marquardt (LM) least-squares algorithm and a Monte Carlo Markov chain (MCMC) Bayesian approach (Metropolis-Hastings sampling). The observational uncertainties of $M_{ISM}$, sSFR/sSFR$_{MS}$, and $M_{*}$ are taken into account. The results of the LM and MCMC fitting are 
\begin{eqnarray}
\notag
LM:\\
\notag
M_{ISM} & =  & 5.85^{+0.26}_{-0.28} \times 10^{9} M_{\odot} \times (1+z)^{1.88^{+0.03}_{-0.03}}  
\\  
\notag
& & \times (sSFR/sSFR_{MS})^{0.20^{+0.02}_{-0.02}} 
\\
& & \times (\frac{M_{*}}{10^{10} M_{\odot}})^{0.24^{+0.02}_{-0.02}} \times \Delta^{0.03^{+0.01}_{-0.01}} \label{eqn:1}
\\
\notag
MCMC:\\
\notag
M_{ISM} & =  & 6.06^{+0.27}_{-0.28} \times 10^{9} M_{\odot} \times (1+z)^{1.83^{+0.03}_{-0.04}}  
\\  
\notag
& & \times (sSFR/sSFR_{MS})^{0.21^{+0.03}_{-0.02}} 
\\
& & \times (\frac{M_{*}}{10^{10} M_{\odot}})^{0.24^{+0.02}_{-0.02}} \times \Delta^{0.03^{+0.01}_{-0.01}}  \label{eqn:2}
\end{eqnarray} 
The LM and MCMC methods both yield consistent results. As expected, the parametric dependence of $M_{ISM}$ on $z$, sSFR/sSFR$_{MS}$, and $M_{*}$ is similar to what was found in \cite{Scoville17}. However, here, we explicitly investigate the potential dependence of $M_{ISM}$ on the environment of galaxies as well (with overdensity of galaxies as a measure of environment). From Equations \ref{eqn:1} and \ref{eqn:2}, we find only a weak dependence of $M_{ISM}$ on the galaxy overdensity, with a power law of the form $M_{ISM}$ $\propto$ $\Delta^{0.03^{+0.01}_{-0.01}}$. 

We also investigate the scaling relations for the SFR (similar to \citealp{Scoville17}) including the overdensity dependence. As noted by \cite{Scoville17}, we impose a linear dependence of the SFR on $M_{ISM}$, with $M_{ISM}$ obtained directly from Equation \ref{eqn:2} rather than returning to the observed $M_{ISM}$ values. This is essential in order to isolate the star-formation efficiency (SFR/$M_{ISM}$) variation with $z$, sSFR/sSFR$_{MS}$, $M_{*}$, and $\Delta$ from the variation of the $M_{ISM}$ with the same parameters. The results of the LM and MCMC fits are
\begin{eqnarray}
\notag
LM:\\
\notag
SFR & =  & 0.38^{+0.01}_{-0.01} M_{\odot} yr^{-1} \times (\frac{M_{ISM}}{10^{9} M_{\odot}})  
\\  
\notag
& & \times (1+z)^{1.07^{+0.02}_{-0.02}} \times (sSFR/sSFR_{MS})^{0.75^{+0.01}_{-0.01}} 
\\
& & \times (\frac{M_{*}}{10^{10} M_{\odot}})^{0.06^{+0.01}_{-0.01}} \times \Delta^{0.012^{+0.004}_{-0.004}} \label{eqn:sfr-lm}
\\
\notag
MCMC:\\
\notag
SFR & =  & 0.35^{+0.02}_{-0.01} M_{\odot} yr^{-1} \times (\frac{M_{ISM}}{10^{9} M_{\odot}})  
\\  
\notag
& & \times (1+z)^{1.11^{+0.04}_{-0.04}} \times (sSFR/sSFR_{MS})^{0.75^{+0.01}_{-0.01}} 
\\
& & \times (\frac{M_{*}}{10^{10} M_{\odot}})^{0.06^{+0.01}_{-0.01}} \times \Delta^{0.004^{+0.007}_{-0.005}} \label{eqn:sfr-mcmc}
\end{eqnarray} 

The LM and MCMC results are consistent with no dependence of the SFR (and subsequently, the star formation efficiency and depletion timescale $\tau$=$M_{ISM}$/SFR) on overdensity values.
 
However, we note that the fitting procedure is only applicable to the range of parameters probed here. A galaxy sample covering a much larger range of overdensity values (e.g., sampled in dense cores of clusters) and other physical parameters are needed to be able to more robustly investigate the dependence of $M_{ISM}$ on galaxy properties, including their environment. It is also worth noting that the overdensity might not be fully independent of the other variables such as the stellar mass used in the fitting procedure. This dependence might be particularly important in the extreme regions of environments and high stellar masses. However, the potential relation between environment and stellar mass is still debated (see e.g., \citealp{Baldry06} and \citealp{Darvish16} versus \citealp{Vonderlinden10}).  

\subsection{General Trends with Environment for the Control Samples}

We also compare galaxy properties in different density quantiles for the controlled samples. Table \ref{table2} summarizes the results in this paper. It contains the average and median gas properties of galaxies in different density quantiles and redshifts. After controlling for $z$, $M_{*}$, and sSFR/sSFR$_{MS}$, we compare the total molecular mass ($M_{ISM}$), gas mass fraction (f$_{gas}$), and depletion timescale ($\tau$) in different environmental bins, as shown in Figure \ref{fig:hist-all}. We see no clear environmental dependence of $M_{ISM}$, f$_{gas}$, and $\tau$ in the local universe and out to $z$ $\sim$ 3.5. We perform K-S tests, and in all cases, the K-S tow-tailed p-value is  $>$ 0.1 (corresponding to a significance of $\lesssim$ 1.6$\sigma$), indicating that it is unlikely that the distributions of $M_{ISM}$, f$_{gas}$, and $\tau$ in different density bins are drawn from different parent distributions.

In the local universe ($\langle z\rangle$ $\sim$ 0.04), the average $\langle M_{ISM}\rangle$,$\langle$f$_{gas}\rangle$, and $\langle \tau \rangle$ in the highest density bin is (3.9$\pm$0.2)$\times$10$^{9}$ $M_{\odot}$, 13.9$\pm$0.7\%, and 1.3$\pm$0.1 Gyr, respectively. These values are reasonably similar to those in the lowest density quantile within the uncertainties ((3.9$\pm$0.2)$\times$10$^{9}$ $M_{\odot}$, 14.5$\pm$0.6\%, and 1.3$\pm$0.1 Gyr). The uncertainties here are the standard error of the mean. For the high-$z$ sample ($\langle z\rangle$ $\sim$ 1.8) and in densest regions, we obtain the average values of $\langle M_{ISM}\rangle$=(1.6$\pm$0.2)$\times$10$^{11}$ $M_{\odot}$, $\langle$f$_{gas}\rangle$=55$\pm$2\%, and $\langle \tau \rangle$=0.8$\pm$0.1 Gyr, in agreement with the results in the lowest density bin ($\langle M_{ISM}\rangle$=(1.4$\pm$0.2)$\times$10$^{11}$ $M_{\odot}$, $\langle$f$_{gas}\rangle$=52$\pm$2\%, and $\langle \tau \rangle$=0.7$\pm$0.1 Gyr; see Table \ref{table2}).

\begin{table*}
\begin{center}
\caption{Fraction of Volume ($f_{V}$) Occupied by and the Global Molecular Mass Density ($\rho$) in Different Environmental Density Quantiles} 
\begin{tiny}
\centering
\begin{tabular}{lcccccccccc}
\hline
\noalign{\smallskip}
Redshift range & $f_{V}(\delta_{1})$\footnote{Fraction of the volume occupied by galaxies located in $\delta_{1}$ quantile.} & $f_{V}(\delta_{2})$ & $f_{V}(\delta_{3})$ & $f_{V}(\delta_{4})$ & $\rho$($\delta_{1}$)\footnote{Global molecular mass density in $\delta_{1}$ quantile.} & $\rho$($\delta_{2}$) & $\rho$($\delta_{3}$) & $\rho$($\delta_{4}$) & $
\rho$(all)\\
& (\%) & (\%) & (\%) & (\%) & (10$^{6}$ $M_{\odot}$Mpc$^{-3}$) & (10$^{6}$ $M_{\odot}$Mpc$^{-3}$) & (10$^{6}$ $M_{\odot}$Mpc$^{-3}$) & (10$^{6}$ $M_{\odot}$Mpc$^{-3}$) & (10$^{6}$ $M_{\odot}$Mpc$^{-3}$)\\
\hline
\\
0.02$-$0.12 & 66.8$\pm$3.6 & 21.5$\pm$2.0 & 9.4$\pm$1.4 & 2.4$\pm$0.2 & 7.7$_{-1.3}^{+1.5}$ & 24.0$_{-4.4}^{+5.0}$ & 55.0$_{-11.8}^{+13.0}$ & 213.8$_{-39.3}^{+44.8}$ & 20.6$_{-3.8}^{+3.0}$\\
\\
0.2$-$0.5 & 72.4$\pm$4.0 & 17.5$\pm$2.1 & 7.2$\pm$1.4 & 2.9$\pm$0.5 & 10.9$_{-2.0}^{+1.9}$ & 44.9$_{-9.7}^{+9.0}$ & 108.8$_{-28.4}^{+26.9}$ & 276.4$_{-72.6}^{+69.0}$  & 31.5$_{-4.6}^{+6.2}$\\
\\
0.5$-$0.8 & 73.6$\pm$4.2 & 16.6$\pm$2.2 & 7.3$\pm$1.4 & 2.6$\pm$0.6 & 11.7$_{-2.4}^{+2.6}$ & 52.1$_{-12.2}^{+13.2}$ & 118.7$_{-32.3}^{+34.2}$ & 338.0$_{-104.2}^{+109.0}$ & 34.5$_{-6.1}^{+7.2}$\\
\\
0.8$-$1.1 & 70.2$\pm$5.2 & 17.4$\pm$2.5 & 8.6$\pm$1.8 & 3.8$\pm$0.9 & 15.4$_{-4.7}^{+5.6}$ & 62.1$_{-20.3}^{+23.8}$ & 125.9$_{-45.3}^{+51.9}$ & 286.1$_{-109.6}^{+123.8}$  & 43.3$_{-12.0}^{+15.2}$\\
\\
1.1$-$1.5 & 61.9$\pm$3.9 & 20.8$\pm$1.4 & 11.6$\pm$1.5 & 5.7$\pm$1.1 & 25.0$_{-4.9}^{+4.8}$ & 74.4$_{-14.5}^{+14.3}$ & 133.4$_{-30.0}^{+29.7}$ & 271.6$_{-71.2}^{+70.7}$  & 61.8$_{-11.3}^{+11.9}$\\
\\
1.5$-$2.0 & 57.7$\pm$3.5 & 22.2$\pm$1.1 & 13.2$\pm$1.4 & 6.9$\pm$1.1 & 28.2$_{-2.9}^{+2.8}$ & 73.2$_{-7.1}^{+6.7}$ & 122.9$_{-16.3}^{+15.8}$ & 235.8$_{-42.1}^{+41.3}$  & 65.1$_{-5.0}^{+6.2}$\\
\\
2.0$-$2.5 & 60.1$\pm$3.9 & 20.5$\pm$1.2 & 12.5$\pm$1.4 & 6.9$\pm$1.3 & 22.1$_{-2.7}^{+2.8}$ & 64.8$_{-7.6}^{+8.0}$ & 106.2$_{-16.1}^{+16.7}$ & 191.3$_{-41.2}^{+42.0}$  & 53.1$_{-5.3}^{+5.9}$\\
\\
2.5$-$3.0 & 61.0$\pm$4.5 & 20.6$\pm$1.7 & 12.1$\pm$1.6 & 6.3$\pm$1.3 & 18.9$_{-6.2}^{+7.4}$ & 56.0$_{-18.4}^{+22.0}$ & 95.8$_{-33.0}^{+38.9}$ & 183.0$_{-68.7}^{+79.2}$  & 46.2$_{-15.8}^{+16.8}$\\
\\
3.0$-$4.0 & 62.1$\pm$4.8 & 19.8$\pm$2.4 & 11.7$\pm$1.5 & 6.5$\pm$0.9 & 11.6$_{-3.6}^{+4.4}$ & 36.4$_{-11.8}^{+14.2}$ & 61.6$_{-20.1}^{+24.2}$ & 111.1$_{-36.7}^{+43.9}$  & 28.8$_{-8.9}^{+9.4}$\\
\\
\hline
\label{table3}
\end{tabular}
\end{tiny}
\end{center}
\end{table*}

\subsection{Environment and Redshift Dependence of the Gas Mass Fraction}

We further investigate the environmental dependence of $M_{ISM}$, f$_{gas}$, and $\tau$ in redshift bins. Note that we control for $z$, $M_{*}$, and sSFR/sSFR$_{MS}$ distributions in different density bins for the redshift-binned data as well. Even after dividing the sample into redshift bins, we still find no significant environmental dependence of $M_{ISM}$, f$_{gas}$, and $\tau$ within the errors.

\begin{figure*}
\centering
\includegraphics[width=7.0in]{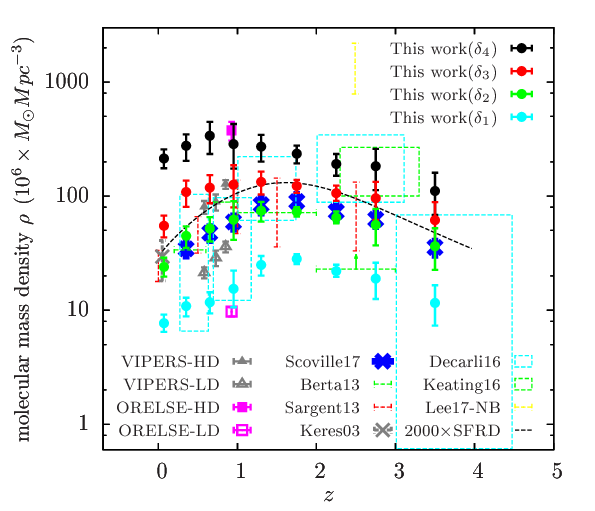}
\caption{Redshift evolution of the molecular mass density $\rho$ in different environmental density quantiles (filled circles) estimated using Equation \ref{eqn:5}. We also show the estimation of $\rho$ by directly using the stellar mass functions in densest (HD) and the least dense (LD) regions from the VIPERS \citep{Davidzon16} and ORELSE \citep{Tomczak17} surveys (only the integral term in Equation \ref{eqn:5} is used). The cluster study of \cite{Lee17} and some observations and simulations for the overall field galaxies are also shown \citep{Keres03,Berta13,Sargent13,Decarli16,Keating16,Scoville17}. At each redshift, denser environments have a higher molecular mass density than the less dense regions, primarily because of a volume effect as they occupy a smaller volume of the total volume of a survey. Our $\rho$ measurements in the lowest density bins peak around $z$ $\sim$ 1-2, and their redshift evolution has a similar shape to that of \cite{Scoville17} and some field studies. The peak in the $\rho$ measurements in the highest density quantiles occurs at $z$ $<$ 1. This differential redshift evolution in low- and high-density environments is likely the result of the growth of the large-scale structure with cosmic time. The SFRD (scaled by a factor of 2000) from Equation (15) of \cite{Madau14} is also shown, depicting a resemblance to the overall evolution of $\rho$.}
\label{fig:rho-z}
\end{figure*}

Figure \ref{fig:fgas-z} shows the redshift evolution of f$_{gas}$ in different density bins, along with cluster \citep{Cybulski16,Lee17,Noble17} and field \citep{Tacconi10,Geach11b,Saintonge11,Bauermeister13,Kirkpatrick14,Cybulski16,Schinnerer16,Scoville17} quantities in the literature. Note that the literature studies have their own sample selections that might be different than ours, and they are just shown for reference. Nonetheless, we find that independent of the environment (within the uncertainties), the gas fraction decreases with decreasing redshift. Moreover, our data points nearly follow the overall decline in f$_{gas}$ with cosmic time.

\subsection{Environment and Redshift Dependence of the Depletion Timescale}

The overall depletion timescale increases with cosmic time, as shown in Figure \ref{fig:tau-z}. However, this increase is almost independent of the environment at each given redshift. For reference, the analytic equations from \cite{Scoville17} (their Tables 2 and 3) and \cite{Genzel15} (using dust-data binned parameters presented in their Table 3, where dust-data refers to gas mass estimates based on a metallicity-dependent dust-to-gas ratio and dust mass estimates using the SFR, and the estimated dust temperature assuming a modified blackbody spectrum and an emissivity index of $\beta$=1.5; see section 2.2 of \citealp{Genzel15}) are shown. We use the median sSFR/sSFR$_{MS}$ and $M_{*}$ of our sample in all density bins in \cite{Scoville17} and \cite{Genzel15} equations. The uncertainties include the model fitting errors and the dispersion in the sSFR/sSFR$_{MS}$ and $M_{*}$ observables (dominated by the latter). Within the uncertainties, our estimated $\tau$ values are consistent with these global trends, regardless of their local environment.

\subsection{Environment and Redshift Dependence of the Molecular Mass Density}
       
We also investigate the redshift evolution and environmental dependence of the global molecular mass density ($\rho$). For this, we rely on the overall molecular mass density at each redshift (see \citealp{Scoville17}) and scale it with the fraction of galaxies in each density quantile (25\%) and the inverse of the fraction of the volume assigned to each density quantile:
\begin{equation} 
\rho(\delta,z)=\frac{1}{4} \frac{1}{f_{V}(\delta,z)} \int M_{ISM}\Phi(M_{ISM},z)dM_{ISM} 
\label{eqn:3}
\end{equation}  
where $\rho(\delta,z)$ is the molecular mass density for the density quantile of $\delta$ and redshift of $z$, $f_{V}(\delta,z)$ is the fraction of the volume assigned to the density quantile $\delta$ at $z$, and $\Phi(M_{ISM},z)$ is the molecular mass function at $z$. $f_{V}(\delta,z)$ can be measured via
\begin{equation} 
f_{V}(\delta,z)=\frac{V(\delta,z)}{V_{total}(z)}, \quad \textrm{where} \quad V(\delta,z)=\sum_{i=1}^{N} \sigma^{-1}_{i} \Delta L_{i}
\label{eqn:4}
\end{equation}
In the above equation, $V(\delta,z)$ is the volume assigned to the density quantile $\delta$ at $z$ and $V_{total}(z)$ is the total volume assigned to all quantiles at that redshift. $V(\delta,z)$ is estimated using all the galaxies $N$(used in estimating the density field) that are located in density quantile $\delta$ at $z$. $\sigma_{i}$ is the projected surface density of the galaxy $i$ (in Mpc$^{-2}$), and $\Delta L_{i}$ is the comoving length (in Mpc) that corresponds to the redshift slice ($\Delta z$) within which the projected surface density of the galaxy $i$ is estimated \footnote{For the local universe sample $\Delta z$=$\pm \Delta v$/c=$\pm$0.0033 \citep{Darvish18} and at high-$z$, $\Delta z$=$\pm$1.5$\sigma_{\Delta z/(1+z)}$, where $\sigma_{\Delta z/(1+z)}$ is the median photo-$z$ uncertainty at each redshift \citep{Darvish17}}.

In order to evaluate the integral in Equation \ref{eqn:3}, we rely on the stellar mass function ($\Phi(M_{*})$) instead of the unknown $\Phi(M_{ISM})$ and the stellar mass dependence of $M_{ISM}$ given in Equation \ref{eqn:2} (also see \citealp{Scoville17}). As seen in Equation \ref{eqn:2}, $M_{ISM}$ also depends on sSFR/sSFR$_{MS}$ and very weakly on the overdensity $\Delta$. Since the majority of active galaxies lie on the main sequence (with starbursts making up only a few percent of the active population; e.g., \citealp{Rodighiero11}), we assume (sSFR/sSFR$_{MS}$)=1 in Equation \ref{eqn:2} when estimating $\rho$. Assuming a 5\% contribution of starburst galaxies and an average (sSFR/sSFR$_{MS}$)=10 for them changes the estimated $\rho$ by only a factor of $\sim$ 1.03. We ignore the $M_{ISM}$ weak dependence on $\Delta$ since changing $\Delta$ by even a factor of 100 results in only a slight $\rho$ increase of a factor of $\sim$ 1.15. We also do not include the contribution of passive galaxies in $\rho$. Given the above assumptions and knowing that $\Phi(M_{ISM},z)dM_{ISM}$=$\Phi(M_{*},z)dM_{*}$, Equation \ref{eqn:3} is simplified to
\begin{equation} 
\rho(\delta,z)=\frac{1}{4} \frac{1}{f_{V}(\delta,z)} \int M_{ISM}(z,M_{*})\Phi(M_{*},z)dM_{*} 
\label{eqn:5}
\end{equation} 
In Equation \ref{eqn:5}, we use the stellar mass function of \cite{Ilbert13} for star-forming galaxies at $z$=0.2-4.0. The integral is performed at $M_{*}$=10$^{10}$ to 10$^{12}$ $M_{\odot}$ because we only used $M_{*}$ $\gtrsim$ 10$^{10}$ $M_{\odot}$ active galaxies in modeling $M_{ISM}(z,sSFR/sSFR_{MS},M_{*},\Delta)$. Using $M_{*}$=10$^{9}$ for the lower limit of integration changes the estimated $\rho$ by a factor of $\sim$ 2.6. For the local universe stellar mass function, we also use that of \cite{Ilbert13} at $z$=0.2-0.5.

The estimated $f_{V}$ and $\rho$ values in different density quantiles and at different redshifts are given in Table \ref{table3}. The uncertainties for $f_{V}$ are estimated by slightly modifying the width of the redshift slices within which projected surface densities are estimated (0.5 $\times$ and 2 $\times$ of the $\Delta z$ adopted in this work; see Section \ref{env}) and recalculating the new $f_{V}$. The maximum difference between the new $f_{V}$ and the original $f_{V}$ is used as this volume-related uncertainty. The uncertainties of $\rho$ have contributions from the uncertainties of $f_{V}$ and those of the parameters in Equation \ref{eqn:2} and the parameters of the stellar mass function. 

Figure \ref{fig:rho-z} shows the estimated molecular mass density as a function of redshift for different density quantiles. We also estimate it by directly using the star-forming stellar mass functions in densest (HD) and the least dense (LD) regions from the VIPERS \citep{Davidzon16} and ORELSE \citep{Tomczak17} surveys (only the integral term in Equation \ref{eqn:5} is used). Note that the definition of the environment and how densest and lowest density regions are defined in these surveys are not necessarily the same as ours. We also show the cluster study of \cite{Lee17} alongside some observational and computational studies for the overall field galaxies \citep{Keres03,Berta13,Sargent13,Decarli16,Keating16,Scoville17}.

At each redshift, denser environments have a higher molecular mass density than the less dense regions. This is mainly due to a volume effect as the denser regions occupy a smaller volume of the total volume of a survey. Our $\rho$ measurements in the lowest density bins ($\delta_{1}$ and $\delta_{2}$) peak around $z$ $\sim$ 1-2, and the redshift evolution has a similar shape to that of \cite{Scoville17} and some field studies as shown in Figure \ref{fig:rho-z}. However, the peak in the $\rho$ measurements in the highest environmental density bins ($\delta_{3}$ and $\delta_{4}$) is shifted to lower redshift of $z$ $<$ 1. This is the result of the growth of the large-scale structure with cosmic time as reflected in the $z$ evolution of the measured fraction of volume occupied by densest regions in Table \ref{table3}. Densest environments occupy a smaller regions of space at lower redshifts than similar environments at higher $z$ due to mergers of structures extended over larger volumes at higher redshifts (see, e.g., \citealp{Chiang13}) and late-time assembly of galaxy groups and clusters. Therefore, the decrease in the global molecular mass at $z$ $\lesssim$ 2 is partly compensated for by the decrease in the volume assigned to dense regions at $z$ $\lesssim$ 2, causing the shift in the peak of the measured $\rho$ in dense regions relative to that of the less dense environments. At $z$=0.65 and in densest regions ($\delta_{4}$), $\rho$ is maximized at 338.0$_{-104.2}^{+109.0} \times$10$^{6}$ $M_{\odot}$Mpc$^{-3}$ and in the lowest density quantile ($\delta_{1}$), it peaks at $z$=1.75 with a value of 28.2$_{-2.9}^{+2.8} \times$10$^{6}$ $M_{\odot}$Mpc$^{-3}$ (Table \ref{table3}).

We also mention that the molecular mass densities between densest and the least dense regions based on \cite{Davidzon16} and \cite{Tomczak17} stellar mass functions coincide with the midpoint of our measurement (between $\delta{1}$ and $\delta_{4}$) and also that of \cite{Scoville17}. This further supports the accuracy of our $\rho$ estimation in different environmental bins.  

Interestingly, the redshift dependence of $\rho$ for the lowest density quantiles (and the overall $\rho$) resembles the shape of the global star-formation rate density (SFRD) of the universe (Madau-Lilly plot; e.g., \citealp{Madau14} and the references therein). For example, our data points in $\delta_{1}$ density quantile rise as (1+$z$)$^{\sim 3.4\pm 1.3}$ and decline as (1+$z$)$^{\sim -2.9}$ with increasing redshift. This is in agreement with rising (1+$z$)$^{2.7}$ and declining (1+$z$)$^{-2.9}$ SFRD from \cite{Madau14} (labeled as 2000$\times$SFRD in Figure \ref{fig:rho-z}). This tempting result indicates that the evolution of the cosmic SFRD is likely governed by the evolution in the cosmic gas content of galaxies, with minimal environmental effects.  

Several studies have found that the shape of the cosmic SFRD in dense environments of clusters and protoclusters is similar to that of the field galaxies or likely has a peak at redshift of $z$ $\gtrsim$ 2, slightly higher than that in the general field (e.g., see \citealp{Clements14,Kato16} and simulations of \citealp{Chiang17}). The peak of the cosmic SFRD in dense environments seems to shift in an opposite sense compared with the shape of the cosmic molecular gas mass density in dense regions found in our study. This apparent difference might be due to a nontrivial and likely an environmental-dependent conversion factor between the global SFRD and that of the molecular mass density. Nonetheless, the details of such study are beyond the scope of this paper.      
 
\section{Discussion} \label{dis}

For $M_{*}$ $\gtrsim$ 10$^{10}$ $M_{\odot}$ star-forming and starburst galaxies, we find no significant evidence for the environmental dependence of their molecular gas mass, gas mass fraction, and depletion timescale since $z$ $\sim$ 3.5. This seems to be at odds with some studies showing that some environmentally driven process such as ram pressure is responsible for stripping the gas and dust content of galaxies in dense environments, resulting in a lower gas mass and gas fraction than their field counterparts (see references in \citealp{Boselli14}).

Ram-pressure stripping is particularly effective in low-mass $\lesssim$ 10$^{9}$ $M_{\odot}$ galaxies (e.g., \citealp{Fillingham16}) as the gravitational bounding force of more massive systems might be strong enough to keep its gas content. However, there is evidence for ram-pressure stripping occurring in more massive galaxies as well (e.g., \citealp{Poggianti17a}). Although observations of the HI deficiencies in cluster galaxies suggest that ram-pressure stripping is significant (e.g., \citealp{Cayatte90,Gavazzi05}), it seems that the molecular gas is less vulnerable to stripping as it is much denser and the observations of the radial distribution of HI and H$_{2}$ gas in nearby galaxies show a much more extended HI than H$_{2}$, making molecular gas more bound to the host galaxy and less prone to stripping. This is supported by the lack of environmental effects on the molecular gas content of galaxies as seen in some studies (e.g., \citealp{Kenney89,Koyama17}). However, others find lower (e.g., \citealp{Fumagalli09}) molecular gas in denser environments, suggesting that ram pressure is still strong enough on stripping the H$_{2}$ gas. The study by \cite{Mok16} finds an excess of molecular gas in denser environments although the difference between H$_{2}$ mass in dense and less dense regions is not significant (within $\sim$ 2$\sigma$ in the best cases), according to their tables 2 and 3.  

Since the estimation of molecular gas for our local universe sample relies on the HI gas, the lack of environmental dependence on the molecular gas content of our low-$z$ sample is also indicative of the environmental independence of the atomic gas at $z$ $\sim$ 0. This might be because of the relatively high mass of our sample galaxies to be affected by ram pressure and/or due to our selection as we are only investigating star-forming and starburst systems.

Selection biases can also play an important role. The gas content of galaxies might depend on redshift, stellar mass, and the sSFR of the galaxy relative to the main sequence (e.g., \citealp{Scoville17}). If $z$, $M_{*}$, or sSFR/sSFR$_{MS}$ distribution of galaxies in dense environments is (intrinsically or due to selections) different than the field samples, this automatically causes a bias that would likely lead to misinterpretation of the results. For example, there might be a correlation between massive systems and denser environments (e.g., \citealp{Bolzonella10,Darvish16}). Many studies also found a lower sSFR in denser environments (especially at lower redshifts) due to a higher fraction of quiescent galaxies and/or a lower SFR of galaxies in denser environments (e.g., \citealp{Kauffmann04,Baldry06,Darvish16}). These automatically bias the denser environment samples to higher $M_{*}$ and lower sSFR/sSFR$_{MS}$ distributions, both leading to lower gas content for galaxies in denser environments. The role of selection biases is recently highlighted by \cite{Koyama17} who found no environmental effects on molecular gas and star-formation efficiency at $z$ $\sim$ 0 after controlling for potential biases.
          
As suggested by some studies, the environment seems to only control the fraction of quiescent/star-forming galaxies \citep{Peng10,Sobral11,Darvish14,Darvish16}; i.e., dense environments increase the likelihood of a galaxy to become quenched, particularly at lower redshifts. However, as long as a galaxy is forming stars, its star-formation activity is not much affected by its host environment. This picture is consistent with our results as we deal with star-forming and starburst galaxies (on the main sequence and above) whose molecular gas content (and the subsequent star-formation activity) is primarily driven by processes (regardless of its physics) other than their local environment. One direct consequence of this scenario is that environmental quenching of galaxies (if any) in terms of gas removal or consumption is a fast process. Otherwise, it would leave its imprint as lower gas content for star-forming and active systems in dense environments.              

We also note that if galaxies in dense environments happen to preferentially populate a region of the 3D $M_{*}$-sSFR/sSFR$_{MS}$-$z$ space that is, for some reason, sparse in galaxies (in less dense bins), then the control sample selection procedure causes a bias, particularly if a large number of them are jettisoned. This is because the procedure (Section \ref{control}) would automatically discard these galaxies as they would not have a representative counterpart in the $M_{*}$-sSFR/sSFR$_{MS}$-$z$ space in lower-density bins. 
 
We also found no evidence for a different depletion timescale in denser environments than the field to $z$ $\sim$ 3.5 consistent with, e.g., \cite{Koyama17} at $z$ $\sim$ 0 and \cite{Noble17} at $z$ $\sim$ 1.6. However, some studies have suggested that denser environments accelerate (decelerate) the consumption of molecular gas, leading to a shorter (longer) $\tau$ in denser regions (e.g., \citealp{Lee17} versus \citealp{Mok16}). However, as noted by \cite{Scoville17}, if the star-forming gas is in self-gravitating giant molecular clouds (GMCs), then the internal structure of the GMCs determines the physics of the star formation, and the star-forming gas does not know if it is in a less or a more massive galaxy, or in our case, if its host galaxy is located in a less or a more dense environment. In other words, the local environment does not seem to influence the amount of $M_{ISM}$ that would go into forming those GMCs.

We note that although our samples cover a large dynamical range of environments, to fully understand the role of extreme environments on the molecular gas content of galaxies, a dedicated survey that targets a large sample of galaxies (e.g. dust-continuum observations with ALMA) in extremely dense cores of confirmed structures at high-$z$ is crucial.

\section{summary} \label{sum}

We investigate the role of environmental density on the molecular gas content of a large sample of $M_{*}$ $\gtrsim$ 10$^{10}$ $M_{\odot}$ star-forming and starburst galaxies to $z$ $\sim$ 3.5. Similar to \cite{Scoville17}, we derive the scaling relations for the molecular mass $M_{ISM}$ and SFR efficiency as a function of redshift ($z$), sSFR relative to the main sequence (sSFR/sSFR$_{MS}$), stellar mass ($M_{*}$), and also galaxy overdensity ($\Delta$). We also investigate the redshift evolution of the global molecular mass density ($\rho$) in different environmental density quantiles. The key results from this paper are listed below.

\begin{enumerate}

\item We find no dependence of the $M_{ISM}$ (subsequently, gas mass fraction f$_{gas}$) and star-formation efficiency (subsequently, depletion time $\tau$) scaling relations on galaxy overdensity. $M_{ISM}$ approximately increases as a power of 0.03 for overdensity $\Delta$. The power term for the star-formation rate efficiency as a function of $\Delta$ is 0.004.
  
\item Similar results are obtained after dividing our sample into four environmental density quantiles ($\delta_{1}$, $\delta_{2}$, $\delta_{3}$, and $\delta_{4}$ from the lowest to highest densities). At $\langle z\rangle$ $\sim$ 0.04, the average $\langle M_{ISM}\rangle$,$\langle$f$_{gas}\rangle$, and $\langle \tau \rangle$ in densest environments is (3.9$\pm$0.2)$\times$10$^{9}$ $M_{\odot}$, 13.9$\pm$0.7\%, and 1.3$\pm$0.1 Gyr, respectively. These values are similar to those in the lowest density quantile ((3.9$\pm$0.2)$\times$10$^{9}$ $M_{\odot}$, 14.5$\pm$0.6\%, and 1.3$\pm$0.1 Gyr). At $\langle z\rangle$ $\sim$ 1.8 and in densest environmental bin, we obtain the average values of $\langle M_{ISM}\rangle$=(1.6$\pm$0.2)$\times$10$^{11}$ $M_{\odot}$, $\langle$f$_{gas}\rangle$=55$\pm$2\%, and $\langle \tau \rangle$=0.8$\pm$0.1 Gyr, in agreement with those in the lowest density quantile ($\langle M_{ISM}\rangle$=(1.4$\pm$0.2)$\times$10$^{11}$ $M_{\odot}$, $\langle$f$_{gas}\rangle$=52$\pm$2\%, and $\langle \tau \rangle$=0.7$\pm$0.1 Gyr). Within the uncertainties, f$_{gas}$ decreases and $\tau$ increases with increasing cosmic time, regardless of the environmental densities (see Table \ref{table2}).

\item At each redshift, denser environments have a higher molecular mass density than the less dense regions. This is mainly due to a volume effect as the denser regions occupy a smaller volume of the total volume of a survey. Our $\rho$ measurements in the lowest density bins ($\delta_{1}$ and $\delta_{2}$) peak around $z$ $\sim$ 1-2, resembling the evolution of the global star-formation rate density. However, the peak in the global $\rho$ is shifted to $z$ $<$ 1 for densest environmental bins ($\delta_{3}$ and $\delta_{4}$). The differential evolution in the peak of the global molecular mass density across different environments is likely the result of the large-scale structure growth with cosmic time. At $z$=0.65 and in densest regions ($\delta_{4}$), $\rho$ is maximized at 338.0$_{-104.2}^{+109.0} \times$10$^{6}$ $M_{\odot}$Mpc$^{-3}$ and in the lowest density quantile ($\delta_{1}$), it peaks at $z$=1.75 with a value of 28.2$_{-2.9}^{+2.8} \times$10$^{6}$ $M_{\odot}$Mpc$^{-3}$ (see Table \ref{table3}). 

\end{enumerate}
   
\section*{acknowledgements}

We are immensely grateful to the anonymous referee for their useful comments and suggestions that significantly improved the quality of this paper. B.D. acknowledges financial support from NASA through the Astrophysics Data Analysis Program (ADAP), grant number NNX12AE20G, and the National Science Foundation, grant number 1716907. B.D. is grateful to Alexandra Pope for her constructive comments and suggestions on the manuscript. B.D. is grateful to Shoubaneh Hemmati, Kirsten Larson, and Iary Davidzon for their thoughtful comments. B.D. wishes to thank Nicola Malavasi for providing their data. T.D.-S. acknowledges support from ALMA-CONICYT project 31130005 and FONDECYT regular project 1151239. This paper makes use of the following ALMA data: ADS/JAO.ALMA 2011.0.00097.S, 2012.1.00076.S, 2012.1.00523.S, 2013.1.00034.S, 2013.1.00111.S, 2015.1.00137.S, 2013.1.00118.S, and 2013.1.00151.S. ALMA is a partnership of ESO (representing its member states), NSF (USA), and NINS (Japan), together with NRC (Canada), NSC and ASIAA (Taiwan), and KASI (Republic of Korea), in cooperation with the Republic of Chile. The Joint ALMA Observatory is operated by ESO, AUI/NRAO, and NAOJ. The National Radio Astronomy Observatory is a facility of the National Science Foundation operated under cooperative agreement by Associated Universities, Inc. The ALFALFA team at Cornell is supported by NSF grants AST-0607007 and AST-1107390 and by grants from the  Brinson Foundation. The Arecibo Observatory is operated by SRI International under a cooperative agreement with the National Science Foundation (AST-1100968), and in alliance with Ana G. Mendez-Universidad Metropolitana, and the Universities Space Research Association. Funding for SDSS-III has been provided by the Alfred P. Sloan Foundation, the Participating Institutions, the National Science Foundation, and the U.S. Department of Energy Office of Science. The SDSS-III web site is http://www.sdss3.org/. SDSS-III is managed by the Astrophysical Research Consortium for the Participating Institutions of the SDSS-III Collaboration including the University of Arizona, the Brazilian Participation Group, Brookhaven National Laboratory, Carnegie Mellon University, University of Florida, the French Participation Group, the German Participation Group, Harvard University, the Instituto de Astrofisica de Canarias, the Michigan State/Notre Dame/JINA Participation Group, Johns Hopkins University, Lawrence Berkeley National Laboratory, Max Planck Institute for Astrophysics, Max Planck Institute for Extraterrestrial Physics, New Mexico State University, New York University, Ohio State University, Pennsylvania State University, University of Portsmouth, Princeton University, the Spanish Participation Group, University of Tokyo, University of Utah, Vanderbilt University, University of Virginia, University of Washington, and Yale University. Based on data products from observations made with ESO Telescopes at the La Silla Paranal Observatory under ESO programme ID 179.A-2005 and on data products produced by TERAPIX and the Cambridge Astronomy Survey Unit on behalf of the UltraVISTA consortium.      

\appendix

\section{A. The Effects of the Control Sample Selection} \label{bias}

The control sample selection presented in Section \ref{control} might lead to some biases in interpreting the environmental dependence of the gas content of star-forming and active galaxies. For example, if galaxies in dense environments happen to preferentially populate a region of the 3D $M_{*}$-sSFR/sSFR$_{MS}$-$z$ space that is, for some reason, sparse in galaxies (in less dense bins), then the control sample selection procedure would lead to a bias, particularly if a large number of them are thrown away in the control sample selection. This is because the procedure would automatically discard these galaxies as they would not have a representative counterpart in the $M_{*}$-sSFR/sSFR$_{MS}$-$z$ space in less dense environments.

\begin{figure*}
\centering
\includegraphics[width=7.0in]{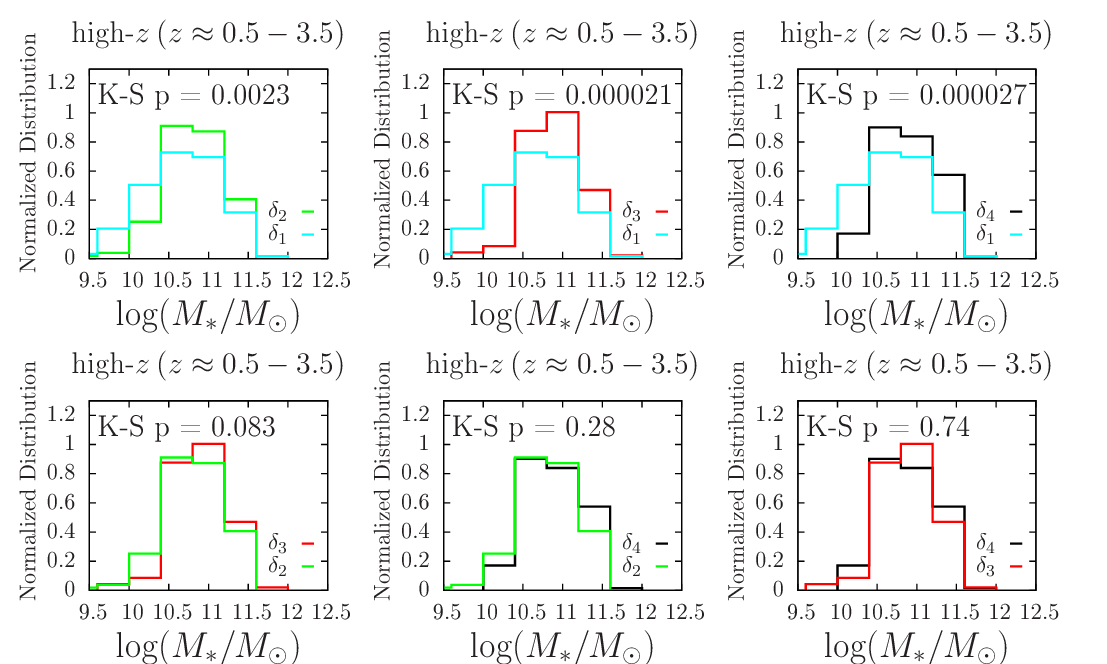}
\caption{Comparison between stellar mass distributions in different environmental density quantiles for our high-$z$ sample, prior to the control sample selection procedure presented in Section \ref{control}. K-S test p-values are also shown.}
\label{fig:dis-Mstar-nc}
\end{figure*}

\begin{figure*}
\centering
\includegraphics[width=7.0in]{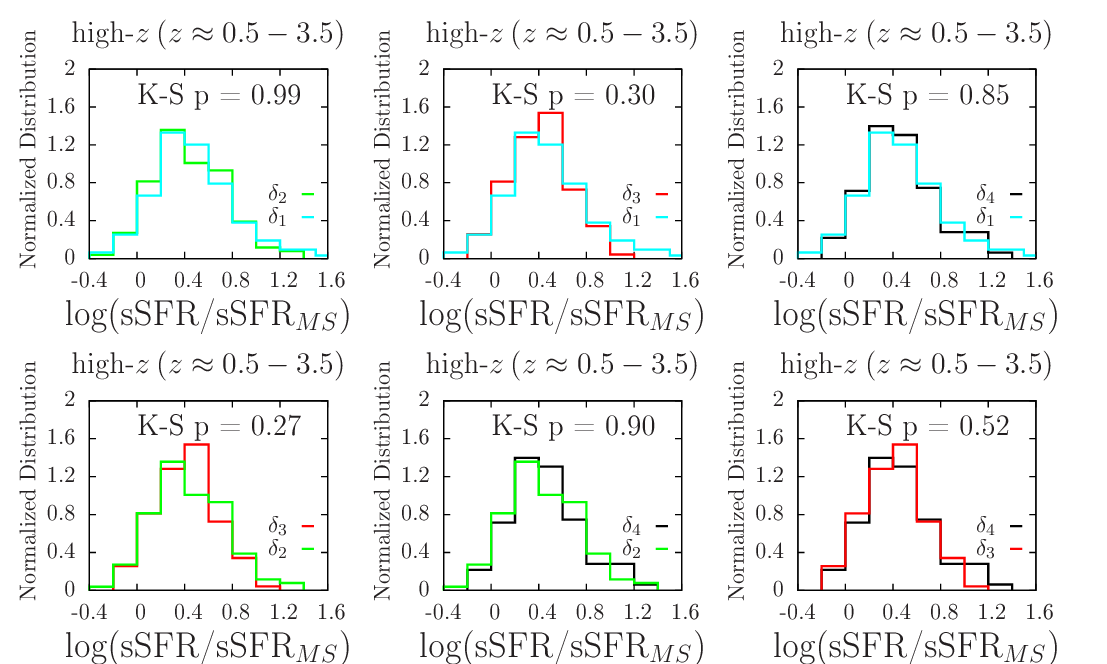}
\caption{Similar to Figure \ref{fig:dis-Mstar-nc} but for the sSFR/sSFR$_{MS}$ comparisons.}
\label{fig:dis-sSFR-nc}
\end{figure*}

\begin{figure*}
\centering
\includegraphics[width=7.0in]{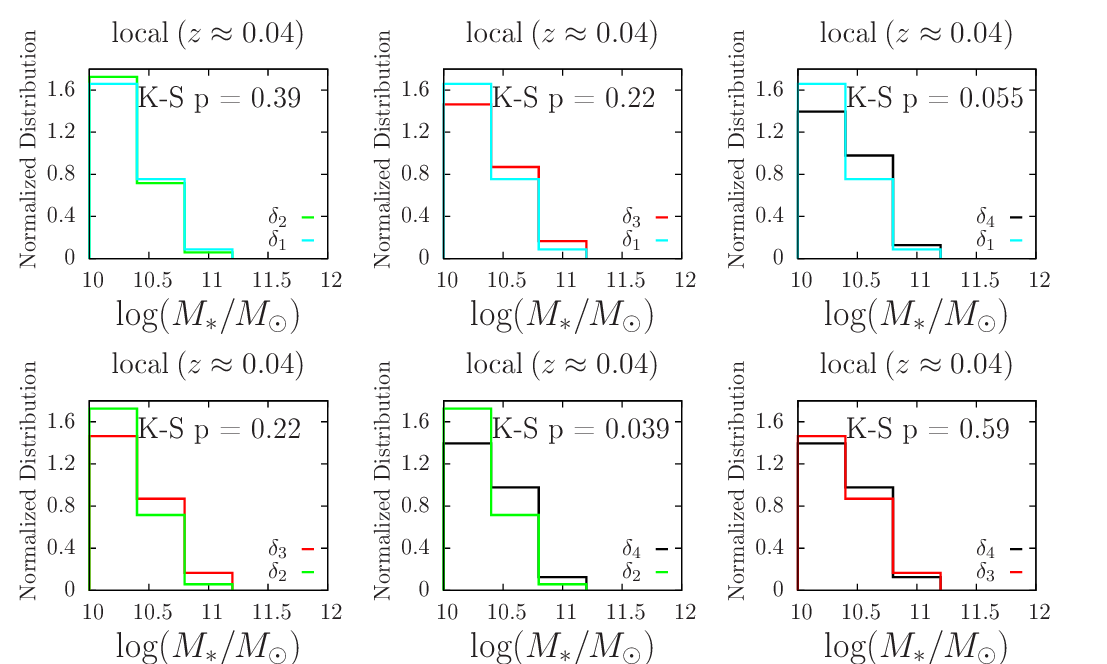}
\caption{Similar to Figure \ref{fig:dis-Mstar-nc} but for the local universe sample.}
\label{fig:dis-Mstar-SDSS-nc}
\end{figure*}

\begin{figure*}
\centering
\includegraphics[width=7.0in]{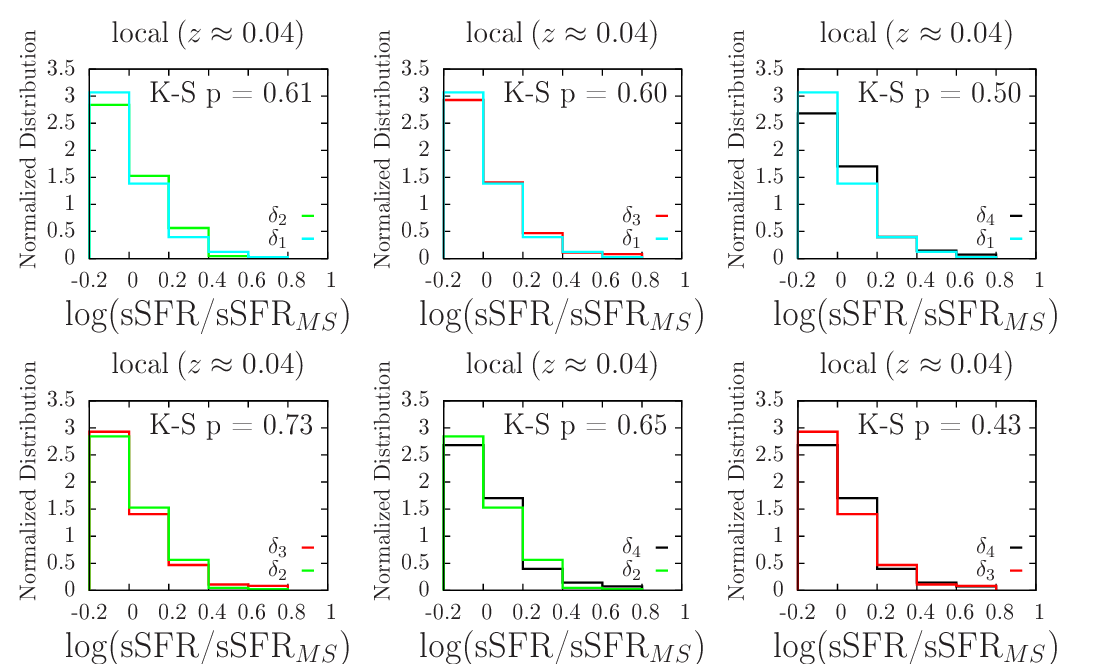}
\caption{Similar to Figure \ref{fig:dis-sSFR-nc} but for the local universe sample.}
\label{fig:dis-sSFR-SDSS-nc}
\end{figure*}

It is also likely that star-forming and starburst galaxies intrinsically present different distributions of stellar mass and sSFR/sSFR$_{MS}$ in different density bins, further biasing the study toward a lack of environmental dependence for the molecular gas. Here, we only control for redshift and present and compare the distribution of stellar mass and sSFR in different density quantiles before controlling them, as shown in Figures \ref{fig:dis-Mstar-nc}, \ref{fig:dis-sSFR-nc}, \ref{fig:dis-Mstar-SDSS-nc}, and \ref{fig:dis-sSFR-SDSS-nc}.

According to Figures \ref{fig:dis-sSFR-nc} and \ref{fig:dis-sSFR-SDSS-nc} and given the K-S p-values, the sSFR/sSFR$_{MS}$  distributions show no significant difference ($\lesssim$ 1.1$\sigma$ in all cases) in different environments for both the local and high-$z$ samples. This is to some extend expected because our sample comprises only active star-forming and starburst galaxies. For these galaxies, a large number of studies including observations and simulations found no significant environmental dependence of their average SFR, sSFR, and the main sequence over a broad redshift range (see, e.g., \citealp{Peng10,Wijesinghe12,Cen14,Darvish14,Darvish15b,Darvish16,Koyama14,Ricciardelli14,
Vogelsberger14,Sobral15,Duivenvoorden16,Hung16}). Even for studies that found an environmental dependence of SFR for star-forming galaxies (e.g., \citealp{Vonderlinden10,Vulcani10,Patel11,Haines13,Tran15,Erfanianfar16,Darvish17,Wang18}), the reduction in the mean SFR for star-forming and active systems in dense environments is often small ($\sim$ 0.1-0.3 dex) and mostly applies to satellite galaxies.

However, stellar mass distributions show statistically significant differences between the least dense and densest environmental bins, particularly for our high-$z$ sample. The largest difference is at the $\sim$ 2.1$\sigma$ (local universe sample) and $\sim$ 4.3$\sigma$ (high-$z$ sample) levels. It is not clear to us whether this is due to an intrinsic difference between stellar masses in different environments or is the result of sample selection (e.g., how the ALMA sources were originally selected for observation). A potential intrinsic correlation between massive galaxies and dense environments is found in several studies (e.g., \citealp{Darvish16,Kawinwanichakij17}) although others found no such relations (e.g., \citealp{Vonderlinden10}). Nonetheless, the presented results using stellar mass controlled samples should be considered with caution. We also note that even without controlling for $M_{*}$, sSFR/sSFR$_{MS}$, and $z$, our multi-variable fits presented in Section \ref{fits} show no overdensity dependence for our sample of $M_{*}$ $\gtrsim$ $10^{10}$ $M_{\odot}$ star-forming and starburst galaxies.     

\bibliographystyle{aasjournal} 
\bibliography{references}

\begin{thebibliography}{}
\expandafter\ifx\csname natexlab\endcsname\relax\def\natexlab#1{#1}\fi
\providecommand{\url}[1]{\href{#1}{#1}}

\bibitem[{{Alam} {et~al.}(2015){Alam}, {Albareti}, {Allende Prieto}, {Anders},
  {Anderson}, {Anderton}, {Andrews}, {Armengaud}, {Aubourg}, {Bailey}, \&
  et~al.}]{Alam15}
{Alam}, S., {Albareti}, F.~D., {Allende Prieto}, C., {et~al.} 2015, \apjs, 219,
  12

\bibitem[{{Aravena} {et~al.}(2012){Aravena}, {Carilli}, {Salvato}, {Tanaka},
  {Lentati}, {Schinnerer}, {Walter}, {Riechers}, {Smolci{\'c}}, {Capak},
  {Aussel}, {Bertoldi}, {Chapman}, {Farrah}, {Finoguenov}, {Le Floc'h}, {Lutz},
  {Magdis}, {Oliver}, {Riguccini}, {Berta}, {Magnelli}, \& {Pozzi}}]{Aravena12}
{Aravena}, M., {Carilli}, C.~L., {Salvato}, M., {et~al.} 2012, \mnras, 426, 258

\bibitem[{{Baldry} {et~al.}(2006){Baldry}, {Balogh}, {Bower}, {Glazebrook},
  {Nichol}, {Bamford}, \& {Budavari}}]{Baldry06}
{Baldry}, I.~K., {Balogh}, M.~L., {Bower}, R.~G., {et~al.} 2006, \mnras, 373,
  469

\bibitem[{{Bauermeister} {et~al.}(2013){Bauermeister}, {Blitz}, {Bolatto},
  {Bureau}, {Leroy}, {Ostriker}, {Teuben}, {Wong}, \&
  {Wright}}]{Bauermeister13}
{Bauermeister}, A., {Blitz}, L., {Bolatto}, A., {et~al.} 2013, \apj, 768, 132

\bibitem[{{Berta} {et~al.}(2013){Berta}, {Lutz}, {Nordon}, {Genzel},
  {Magnelli}, {Popesso}, {Rosario}, {Saintonge}, {Wuyts}, \&
  {Tacconi}}]{Berta13}
{Berta}, S., {Lutz}, D., {Nordon}, R., {et~al.} 2013, \aap, 555, L8

\bibitem[{{Bolzonella} {et~al.}(2010){Bolzonella}, {Kova{\v c}}, {Pozzetti},
  {Zucca}, {Cucciati}, {Lilly}, {Peng}, {Iovino}, {Zamorani}, {Vergani},
  {Tasca}, {Lamareille}, {Oesch}, {Caputi}, {Kampczyk}, {Bardelli}, {Maier},
  {Abbas}, {Knobel}, {Scodeggio}, {Carollo}, {Contini}, {Kneib}, {Le
  F{\`e}vre}, {Mainieri}, {Renzini}, {Bongiorno}, {Coppa}, {de la Torre}, {de
  Ravel}, {Franzetti}, {Garilli}, {Le Borgne}, {Le Brun}, {Mignoli},
  {Pell{\'o}}, {Perez-Montero}, {Ricciardelli}, {Silverman}, {Tanaka},
  {Tresse}, {Bottini}, {Cappi}, {Cassata}, {Cimatti}, {Guzzo}, {Koekemoer},
  {Leauthaud}, {Maccagni}, {Marinoni}, {McCracken}, {Memeo}, {Meneux},
  {Porciani}, {Scaramella}, {Aussel}, {Capak}, {Halliday}, {Ilbert},
  {Kartaltepe}, {Salvato}, {Sanders}, {Scarlata}, {Scoville}, {Taniguchi}, \&
  {Thompson}}]{Bolzonella10}
{Bolzonella}, M., {Kova{\v c}}, K., {Pozzetti}, L., {et~al.} 2010, \aap, 524,
  A76

\bibitem[{{Boselli} {et~al.}(2014){Boselli}, {Cortese}, {Boquien}, {Boissier},
  {Catinella}, {Gavazzi}, {Lagos}, \& {Saintonge}}]{Boselli14c}
{Boselli}, A., {Cortese}, L., {Boquien}, M., {et~al.} 2014, \aap, 564, A67

\bibitem[{{Boselli} \& {Gavazzi}(2014)}]{Boselli14}
{Boselli}, A., \& {Gavazzi}, G. 2014, \aapr, 22, 74

\bibitem[{{Brown} {et~al.}(2017){Brown}, {Catinella}, {Cortese}, {Lagos},
  {Dav{\'e}}, {Kilborn}, {Haynes}, {Giovanelli}, \& {Rafieferantsoa}}]{Brown17}
{Brown}, T., {Catinella}, B., {Cortese}, L., {et~al.} 2017, \mnras, 466, 1275

\bibitem[{{Carilli} \& {Walter}(2013)}]{Carilli13}
{Carilli}, C.~L., \& {Walter}, F. 2013, \araa, 51, 105

\bibitem[{{Catinella} {et~al.}(2013){Catinella}, {Schiminovich}, {Cortese},
  {Fabello}, {Hummels}, {Moran}, {Lemonias}, {Cooper}, {Wu}, {Heckman}, \&
  {Wang}}]{Catinella13}
{Catinella}, B., {Schiminovich}, D., {Cortese}, L., {et~al.} 2013, \mnras, 436,
  34

\bibitem[{{Cayatte} {et~al.}(1990){Cayatte}, {van Gorkom}, {Balkowski}, \&
  {Kotanyi}}]{Cayatte90}
{Cayatte}, V., {van Gorkom}, J.~H., {Balkowski}, C., \& {Kotanyi}, C. 1990,
  \aj, 100, 604

\bibitem[{{Cen}(2014)}]{Cen14}
{Cen}, R. 2014, \apj, 781, 38

\bibitem[{{Chabrier}(2003)}]{Chabrier03}
{Chabrier}, G. 2003, \pasp, 115, 763

\bibitem[{{Chiang} {et~al.}(2013){Chiang}, {Overzier}, \&
  {Gebhardt}}]{Chiang13}
{Chiang}, Y.-K., {Overzier}, R., \& {Gebhardt}, K. 2013, \apj, 779, 127

\bibitem[{{Chiang} {et~al.}(2017){Chiang}, {Overzier}, {Gebhardt}, \&
  {Henriques}}]{Chiang17}
{Chiang}, Y.-K., {Overzier}, R.~A., {Gebhardt}, K., \& {Henriques}, B. 2017,
  \apjl, 844, L23

\bibitem[{{Clements} {et~al.}(2014){Clements}, {Braglia}, {Hyde},
  {P{\'e}rez-Fournon}, {Bock}, {Cava}, {Chapman}, {Conley}, {Cooray}, {Farrah},
  {Gonz{\'a}lez Solares}, {Marchetti}, {Marsden}, {Oliver}, {Roseboom},
  {Schulz}, {Smith}, {Vaccari}, {Vieira}, {Viero}, {Wang}, {Wardlow}, {Zemcov},
  \& {de Zotti}}]{Clements14}
{Clements}, D.~L., {Braglia}, F.~G., {Hyde}, A.~K., {et~al.} 2014, \mnras, 439,
  1193

\bibitem[{{Corbelli} {et~al.}(2012){Corbelli}, {Bianchi}, {Cortese},
  {Giovanardi}, {Magrini}, {Pappalardo}, {Boselli}, {Bendo}, {Davies},
  {Grossi}, {Madden}, {Smith}, {Vlahakis}, {Auld}, {Baes}, {De Looze}, {Fritz},
  {Pohlen}, \& {Verstappen}}]{Corbelli12}
{Corbelli}, E., {Bianchi}, S., {Cortese}, L., {et~al.} 2012, \aap, 542, A32

\bibitem[{{Cortese} {et~al.}(2011){Cortese}, {Catinella}, {Boissier},
  {Boselli}, \& {Heinis}}]{Cortese11}
{Cortese}, L., {Catinella}, B., {Boissier}, S., {Boselli}, A., \& {Heinis}, S.
  2011, \mnras, 415, 1797

\bibitem[{{Cybulski} {et~al.}(2016){Cybulski}, {Yun}, {Erickson}, {De la Luz},
  {Narayanan}, {Monta{\~n}a}, {S{\'a}nchez}, {Zavala}, {Zeballos}, {Chung},
  {Fern{\'a}ndez}, {van Gorkom}, {Haines}, {Jaff{\'e}}, {Montero-Casta{\~n}o},
  {Poggianti}, {Verheijen}, {Yoon}, {Deshev}, {Harrington}, {Hughes},
  {Morrison}, {Schloerb}, \& {Velazquez}}]{Cybulski16}
{Cybulski}, R., {Yun}, M.~S., {Erickson}, N., {et~al.} 2016, \mnras, 459, 3287

\bibitem[{{Daddi} {et~al.}(2010){Daddi}, {Bournaud}, {Walter}, {Dannerbauer},
  {Carilli}, {Dickinson}, {Elbaz}, {Morrison}, {Riechers}, {Onodera}, {Salmi},
  {Krips}, \& {Stern}}]{Daddi10}
{Daddi}, E., {Bournaud}, F., {Walter}, F., {et~al.} 2010, \apj, 713, 686

\bibitem[{{Darvish} {et~al.}(2018){Darvish}, {Martin}, {Gon{\c c}alves},
  {Mobasher}, {Scoville}, \& {Sobral}}]{Darvish18}
{Darvish}, B., {Martin}, C., {Gon{\c c}alves}, T.~S., {et~al.} 2018, \apj, 853,
  155

\bibitem[{{Darvish} {et~al.}(2017){Darvish}, {Mobasher}, {Martin}, {Sobral},
  {Scoville}, {Stroe}, {Hemmati}, \& {Kartaltepe}}]{Darvish17}
{Darvish}, B., {Mobasher}, B., {Martin}, D.~C., {et~al.} 2017, \apj, 837, 16

\bibitem[{{Darvish} {et~al.}(2015{\natexlab{a}}){Darvish}, {Mobasher},
  {Sobral}, {Hemmati}, {Nayyeri}, \& {Shivaei}}]{Darvish15b}
{Darvish}, B., {Mobasher}, B., {Sobral}, D., {et~al.} 2015{\natexlab{a}}, \apj,
  814, 84

\bibitem[{{Darvish} {et~al.}(2016){Darvish}, {Mobasher}, {Sobral}, {Rettura},
  {Scoville}, {Faisst}, \& {Capak}}]{Darvish16}
---. 2016, \apj, 825, 113

\bibitem[{{Darvish} {et~al.}(2015{\natexlab{b}}){Darvish}, {Mobasher},
  {Sobral}, {Scoville}, \& {Aragon-Calvo}}]{Darvish15a}
{Darvish}, B., {Mobasher}, B., {Sobral}, D., {Scoville}, N., \& {Aragon-Calvo},
  M. 2015{\natexlab{b}}, \apj, 805, 121

\bibitem[{{Darvish} {et~al.}(2014){Darvish}, {Sobral}, {Mobasher}, {Scoville},
  {Best}, {Sales}, \& {Smail}}]{Darvish14}
{Darvish}, B., {Sobral}, D., {Mobasher}, B., {et~al.} 2014, \apj, 796, 51

\bibitem[{{Davidzon} {et~al.}(2016){Davidzon}, {Cucciati}, {Bolzonella}, {De
  Lucia}, {Zamorani}, {Arnouts}, {Moutard}, {Ilbert}, {Garilli}, {Scodeggio},
  {Guzzo}, {Abbas}, {Adami}, {Bel}, {Bottini}, {Branchini}, {Cappi}, {Coupon},
  {de la Torre}, {Di Porto}, {Fritz}, {Franzetti}, {Fumana}, {Granett},
  {Guennou}, {Iovino}, {Krywult}, {Le Brun}, {Le F{\`e}vre}, {Maccagni},
  {Ma{\l}ek}, {Marulli}, {McCracken}, {Mellier}, {Moscardini}, {Polletta},
  {Pollo}, {Tasca}, {Tojeiro}, {Vergani}, \& {Zanichelli}}]{Davidzon16}
{Davidzon}, I., {Cucciati}, O., {Bolzonella}, M., {et~al.} 2016, \aap, 586, A23

\bibitem[{{Decarli} {et~al.}(2016){Decarli}, {Walter}, {Aravena}, {Carilli},
  {Bouwens}, {da Cunha}, {Daddi}, {Ivison}, {Popping}, {Riechers}, {Smail},
  {Swinbank}, {Weiss}, {Anguita}, {Assef}, {Bauer}, {Bell}, {Bertoldi},
  {Chapman}, {Colina}, {Cortes}, {Cox}, {Dickinson}, {Elbaz},
  {G{\'o}nzalez-L{\'o}pez}, {Ibar}, {Infante}, {Hodge}, {Karim}, {Le Fevre},
  {Magnelli}, {Neri}, {Oesch}, {Ota}, {Rix}, {Sargent}, {Sheth}, {van der Wel},
  {van der Werf}, \& {Wagg}}]{Decarli16}
{Decarli}, R., {Walter}, F., {Aravena}, M., {et~al.} 2016, \apj, 833, 69

\bibitem[{{Dressler}(1980)}]{Dressler80}
{Dressler}, A. 1980, \apj, 236, 351

\bibitem[{{Duivenvoorden} {et~al.}(2016){Duivenvoorden}, {Oliver}, {Buat},
  {Darvish}, {Efstathiou}, {Farrah}, {Griffin}, {Hurley}, {Ibar}, {Jarvis},
  {Papadopoulos}, {Sargent}, {Scott}, {Scudder}, {Symeonidis}, {Vaccari},
  {Viero}, \& {Wang}}]{Duivenvoorden16}
{Duivenvoorden}, S., {Oliver}, S., {Buat}, V., {et~al.} 2016, \mnras, 462, 277

\bibitem[{{Erfanianfar} {et~al.}(2016){Erfanianfar}, {Popesso}, {Finoguenov},
  {Wilman}, {Wuyts}, {Biviano}, {Salvato}, {Mirkazemi}, {Morselli}, {Ziparo},
  {Nandra}, {Lutz}, {Elbaz}, {Dickinson}, {Tanaka}, {Altieri}, {Aussel},
  {Bauer}, {Berta}, {Bielby}, {Brandt}, {Cappelluti}, {Cimatti}, {Cooper},
  {Fadda}, {Ilbert}, {Le Floch}, {Magnelli}, {Mulchaey}, {Nordon}, {Newman},
  {Poglitsch}, \& {Pozzi}}]{Erfanianfar16}
{Erfanianfar}, G., {Popesso}, P., {Finoguenov}, A., {et~al.} 2016, \mnras, 455,
  2839

\bibitem[{{Fillingham} {et~al.}(2016){Fillingham}, {Cooper}, {Pace},
  {Boylan-Kolchin}, {Bullock}, {Garrison-Kimmel}, \& {Wheeler}}]{Fillingham16}
{Fillingham}, S.~P., {Cooper}, M.~C., {Pace}, A.~B., {et~al.} 2016, \mnras,
  463, 1916

\bibitem[{{Finoguenov} {et~al.}(2007){Finoguenov}, {Guzzo}, {Hasinger},
  {Scoville}, {Aussel}, {B{\"o}hringer}, {Brusa}, {Capak}, {Cappelluti},
  {Comastri}, {Giodini}, {Griffiths}, {Impey}, {Koekemoer}, {Kneib},
  {Leauthaud}, {Le F{\`e}vre}, {Lilly}, {Mainieri}, {Massey}, {McCracken},
  {Mobasher}, {Murayama}, {Peacock}, {Sakelliou}, {Schinnerer}, {Silverman},
  {Smol{\v c}i{\'c}}, {Taniguchi}, {Tasca}, {Taylor}, {Trump}, \&
  {Zamorani}}]{Finoguenov07}
{Finoguenov}, A., {Guzzo}, L., {Hasinger}, G., {et~al.} 2007, \apjs, 172, 182

\bibitem[{{Fumagalli} {et~al.}(2009){Fumagalli}, {Krumholz}, {Prochaska},
  {Gavazzi}, \& {Boselli}}]{Fumagalli09}
{Fumagalli}, M., {Krumholz}, M.~R., {Prochaska}, J.~X., {Gavazzi}, G., \&
  {Boselli}, A. 2009, \apj, 697, 1811

\bibitem[{{Gavazzi} {et~al.}(2005){Gavazzi}, {Boselli}, {van Driel}, \&
  {O'Neil}}]{Gavazzi05}
{Gavazzi}, G., {Boselli}, A., {van Driel}, W., \& {O'Neil}, K. 2005, \aap, 429,
  439

\bibitem[{{Geach} {et~al.}(2011){Geach}, {Smail}, {Moran}, {MacArthur},
  {Lagos}, \& {Edge}}]{Geach11b}
{Geach}, J.~E., {Smail}, I., {Moran}, S.~M., {et~al.} 2011, \apjl, 730, L19

\bibitem[{{Genzel} {et~al.}(2015){Genzel}, {Tacconi}, {Lutz}, {Saintonge},
  {Berta}, {Magnelli}, {Combes}, {Garc{\'{\i}}a-Burillo}, {Neri}, {Bolatto},
  {Contini}, {Lilly}, {Boissier}, {Boone}, {Bouch{\'e}}, {Bournaud}, {Burkert},
  {Carollo}, {Colina}, {Cooper}, {Cox}, {Feruglio}, {F{\"o}rster Schreiber},
  {Freundlich}, {Gracia-Carpio}, {Juneau}, {Kovac}, {Lippa}, {Naab}, {Salome},
  {Renzini}, {Sternberg}, {Walter}, {Weiner}, {Weiss}, \& {Wuyts}}]{Genzel15}
{Genzel}, R., {Tacconi}, L.~J., {Lutz}, D., {et~al.} 2015, \apj, 800, 20

\bibitem[{{Haines} {et~al.}(2013){Haines}, {Pereira}, {Smith}, {Egami},
  {Sanderson}, {Babul}, {Finoguenov}, {Merluzzi}, {Busarello}, {Rawle}, \&
  {Okabe}}]{Haines13}
{Haines}, C.~P., {Pereira}, M.~J., {Smith}, G.~P., {et~al.} 2013, \apj, 775,
  126

\bibitem[{{Hayashi} {et~al.}(2017){Hayashi}, {Kodama}, {Kohno}, {Yamaguchi},
  {Tadaki}, {Hatsukade}, {Koyama}, {Shimakawa}, {Tamura}, \&
  {Suzuki}}]{Hayashi17}
{Hayashi}, M., {Kodama}, T., {Kohno}, K., {et~al.} 2017, \apjl, 841, L21

\bibitem[{{Hayashi} {et~al.}(2018){Hayashi}, {Tadaki}, {Kodama}, {Kohno},
  {Yamaguchi}, {Hatsukade}, {Koyama}, {Shimakawa}, {Tamura}, \&
  {Suzuki}}]{Hayashi18}
{Hayashi}, M., {Tadaki}, K.-i., {Kodama}, T., {et~al.} 2018, \apj, 856, 118

\bibitem[{{Haynes} {et~al.}(2011){Haynes}, {Giovanelli}, {Martin}, {Hess},
  {Saintonge}, {Adams}, {Hallenbeck}, {Hoffman}, {Huang}, {Kent}, {Koopmann},
  {Papastergis}, {Stierwalt}, {Balonek}, {Craig}, {Higdon}, {Kornreich},
  {Miller}, {O'Donoghue}, {Olowin}, {Rosenberg}, {Spekkens}, {Troischt}, \&
  {Wilcots}}]{Haynes11}
{Haynes}, M.~P., {Giovanelli}, R., {Martin}, A.~M., {et~al.} 2011, \aj, 142,
  170

\bibitem[{{Hung} {et~al.}(2016){Hung}, {Casey}, {Chiang}, {Capak}, {Cowley},
  {Darvish}, {Kacprzak}, {Kova{\v c}}, {Lilly}, {Nanayakkara}, {Spitler},
  {Tran}, \& {Yuan}}]{Hung16}
{Hung}, C.-L., {Casey}, C.~M., {Chiang}, Y.-K., {et~al.} 2016, \apj, 826, 130

\bibitem[{{Ilbert} {et~al.}(2013){Ilbert}, {McCracken}, {Le F{\`e}vre},
  {Capak}, {Dunlop}, {Karim}, {Renzini}, {Caputi}, {Boissier}, {Arnouts},
  {Aussel}, {Comparat}, {Guo}, {Hudelot}, {Kartaltepe}, {Kneib}, {Krogager},
  {Le Floc'h}, {Lilly}, {Mellier}, {Milvang-Jensen}, {Moutard}, {Onodera},
  {Richard}, {Salvato}, {Sanders}, {Scoville}, {Silverman}, {Taniguchi},
  {Tasca}, {Thomas}, {Toft}, {Tresse}, {Vergani}, {Wolk}, \& {Zirm}}]{Ilbert13}
{Ilbert}, O., {McCracken}, H.~J., {Le F{\`e}vre}, O., {et~al.} 2013, \aap, 556,
  A55

\bibitem[{{Jablonka} {et~al.}(2013){Jablonka}, {Combes}, {Rines}, {Finn}, \&
  {Welch}}]{Jablonka13}
{Jablonka}, P., {Combes}, F., {Rines}, K., {Finn}, R., \& {Welch}, T. 2013,
  \aap, 557, A103

\bibitem[{{Jaff{\'e}} {et~al.}(2015){Jaff{\'e}}, {Smith}, {Candlish},
  {Poggianti}, {Sheen}, \& {Verheijen}}]{Jaffe15}
{Jaff{\'e}}, Y.~L., {Smith}, R., {Candlish}, G.~N., {et~al.} 2015, \mnras, 448,
  1715

\bibitem[{{Kato} {et~al.}(2016){Kato}, {Matsuda}, {Smail}, {Swinbank},
  {Hatsukade}, {Umehata}, {Tanaka}, {Saito}, {Iono}, {Tamura}, {Kohno}, {Erb},
  {Lehmer}, {Geach}, {Steidel}, {Alexander}, {Yamada}, \& {Hayashino}}]{Kato16}
{Kato}, Y., {Matsuda}, Y., {Smail}, I., {et~al.} 2016, \mnras, 460, 3861

\bibitem[{{Kauffmann} {et~al.}(2004){Kauffmann}, {White}, {Heckman},
  {M{\'e}nard}, {Brinchmann}, {Charlot}, {Tremonti}, \&
  {Brinkmann}}]{Kauffmann04}
{Kauffmann}, G., {White}, S.~D.~M., {Heckman}, T.~M., {et~al.} 2004, \mnras,
  353, 713

\bibitem[{{Kauffmann} {et~al.}(2003){Kauffmann}, {Heckman}, {White}, {Charlot},
  {Tremonti}, {Brinchmann}, {Bruzual}, {Peng}, {Seibert}, {Bernardi},
  {Blanton}, {Brinkmann}, {Castander}, {Cs{\'a}bai}, {Fukugita}, {Ivezic},
  {Munn}, {Nichol}, {Padmanabhan}, {Thakar}, {Weinberg}, \&
  {York}}]{Kauffmann03}
{Kauffmann}, G., {Heckman}, T.~M., {White}, S.~D.~M., {et~al.} 2003, \mnras,
  341, 33

\bibitem[{{Kawinwanichakij} {et~al.}(2017){Kawinwanichakij}, {Papovich},
  {Quadri}, {Glazebrook}, {Kacprzak}, {Allen}, {Bell}, {Croton}, {Dekel},
  {Ferguson}, {Forrest}, {Grogin}, {Guo}, {Kocevski}, {Koekemoer}, {Labb{\'e}},
  {Lucas}, {Nanayakkara}, {Spitler}, {Straatman}, {Tran}, {Tomczak}, \& {van
  Dokkum}}]{Kawinwanichakij17}
{Kawinwanichakij}, L., {Papovich}, C., {Quadri}, R.~F., {et~al.} 2017, \apj,
  847, 134

\bibitem[{{Keating} {et~al.}(2016){Keating}, {Marrone}, {Bower}, {Leitch},
  {Carlstrom}, \& {DeBoer}}]{Keating16}
{Keating}, G.~K., {Marrone}, D.~P., {Bower}, G.~C., {et~al.} 2016, \apj, 830,
  34

\bibitem[{{Kenney} \& {Young}(1989)}]{Kenney89}
{Kenney}, J.~D.~P., \& {Young}, J.~S. 1989, \apj, 344, 171

\bibitem[{{Keres} {et~al.}(2003){Keres}, {Yun}, \& {Young}}]{Keres03}
{Keres}, D., {Yun}, M.~S., \& {Young}, J.~S. 2003, \apj, 582, 659

\bibitem[{{Kirkpatrick} {et~al.}(2014){Kirkpatrick}, {Pope}, {Aretxaga},
  {Armus}, {Calzetti}, {Helou}, {Monta{\~n}a}, {Narayanan}, {Schloerb}, {Shi},
  {Vega}, \& {Yun}}]{Kirkpatrick14}
{Kirkpatrick}, A., {Pope}, A., {Aretxaga}, I., {et~al.} 2014, \apj, 796, 135

\bibitem[{{Koyama} {et~al.}(2017){Koyama}, {Koyama}, {Yamashita},
  {Morokuma-Matsui}, {Matsuhara}, {Nakagawa}, {Hayashi}, {Kodama}, {Shimakawa},
  {Suzuki}, {Tadaki}, {Tanaka}, \& {Yamamoto}}]{Koyama17}
{Koyama}, S., {Koyama}, Y., {Yamashita}, T., {et~al.} 2017, \apj, 847, 137

\bibitem[{{Koyama} {et~al.}(2014){Koyama}, {Kodama}, {Tadaki}, {Hayashi},
  {Tanaka}, \& {Shimakawa}}]{Koyama14}
{Koyama}, Y., {Kodama}, T., {Tadaki}, K.-i., {et~al.} 2014, \apj, 789, 18

\bibitem[{{Laigle} {et~al.}(2016){Laigle}, {McCracken}, {Ilbert}, {Hsieh},
  {Davidzon}, {Capak}, {Hasinger}, {Silverman}, {Pichon}, {Coupon}, {Aussel},
  {Le Borgne}, {Caputi}, {Cassata}, {Chang}, {Civano}, {Dunlop}, {Fynbo},
  {Kartaltepe}, {Koekemoer}, {Le F{\`e}vre}, {Le Floc'h}, {Leauthaud}, {Lilly},
  {Lin}, {Marchesi}, {Milvang-Jensen}, {Salvato}, {Sanders}, {Scoville},
  {Smolcic}, {Stockmann}, {Taniguchi}, {Tasca}, {Toft}, {Vaccari}, \&
  {Zabl}}]{Laigle16}
{Laigle}, C., {McCracken}, H.~J., {Ilbert}, O., {et~al.} 2016, \apjs, 224, 24

\bibitem[{{Lavezzi} \& {Dickey}(1998)}]{Lavezzi98}
{Lavezzi}, T.~E., \& {Dickey}, J.~M. 1998, \aj, 115, 405

\bibitem[{{Lee} {et~al.}(2017){Lee}, {Tanaka}, {Kawabe}, {Kohno}, {Kodama},
  {Kajisawa}, {Yun}, {Nakanishi}, {Iono}, {Tamura}, {Hatsukade}, {Umehata},
  {Saito}, {Izumi}, {Aretxaga}, {Tadaki}, {Zeballos}, {Ikarashi}, {Wilson},
  {Hughes}, \& {Ivison}}]{Lee17}
{Lee}, M.~M., {Tanaka}, I., {Kawabe}, R., {et~al.} 2017, \apj, 842, 55

\bibitem[{{Lee} {et~al.}(2013){Lee}, {Sanders}, {Casey}, {Scoville}, {Hung},
  {Le Floc'h}, {Ilbert}, {Aussel}, {Capak}, {Kartaltepe}, {Roseboom},
  {Salvato}, {Aravena}, {Berta}, {Bock}, {Oliver}, {Riguccini}, \&
  {Symeonidis}}]{Lee13}
{Lee}, N., {Sanders}, D.~B., {Casey}, C.~M., {et~al.} 2013, \apj, 778, 131

\bibitem[{{Lee} {et~al.}(2015){Lee}, {Sanders}, {Casey}, {Toft}, {Scoville},
  {Hung}, {Le Floc'h}, {Ilbert}, {Zahid}, {Aussel}, {Capak}, {Kartaltepe},
  {Kewley}, {Li}, {Schawinski}, {Sheth}, \& {Xiao}}]{Lee15a}
---. 2015, \apj, 801, 80

\bibitem[{{Madau} \& {Dickinson}(2014)}]{Madau14}
{Madau}, P., \& {Dickinson}, M. 2014, \araa, 52, 415

\bibitem[{{Mok} {et~al.}(2016){Mok}, {Wilson}, {Golding}, {Warren}, {Israel},
  {Serjeant}, {Knapen}, {S{\'a}nchez-Gallego}, {Barmby}, {Bendo}, {Rosolowsky},
  \& {van der Werf}}]{Mok16}
{Mok}, A., {Wilson}, C.~D., {Golding}, J., {et~al.} 2016, \mnras, 456, 4384

\bibitem[{{Noble} {et~al.}(2017){Noble}, {McDonald}, {Muzzin}, {Nantais},
  {Rudnick}, {van Kampen}, {Webb}, {Wilson}, {Yee}, {Boone}, {Cooper},
  {DeGroot}, {Delahaye}, {Demarco}, {Foltz}, {Hayden}, {Lidman},
  {Manilla-Robles}, \& {Perlmutter}}]{Noble17}
{Noble}, A.~G., {McDonald}, M., {Muzzin}, A., {et~al.} 2017, \apjl, 842, L21

\bibitem[{{Obreschkow} \& {Rawlings}(2009)}]{Obreschkow09}
{Obreschkow}, D., \& {Rawlings}, S. 2009, \mnras, 394, 1857

\bibitem[{{Owers} {et~al.}(2012){Owers}, {Couch}, {Nulsen}, \&
  {Randall}}]{Owers12}
{Owers}, M.~S., {Couch}, W.~J., {Nulsen}, P.~E.~J., \& {Randall}, S.~W. 2012,
  \apjl, 750, L23

\bibitem[{{Patel} {et~al.}(2011){Patel}, {Kelson}, {Holden}, {Franx}, \&
  {Illingworth}}]{Patel11}
{Patel}, S.~G., {Kelson}, D.~D., {Holden}, B.~P., {Franx}, M., \&
  {Illingworth}, G.~D. 2011, \apj, 735, 53

\bibitem[{{Peng} {et~al.}(2010){Peng}, {Lilly}, {Kova{\v c}}, {Bolzonella},
  {Pozzetti}, {Renzini}, {Zamorani}, {Ilbert}, {Knobel}, {Iovino}, {Maier},
  {Cucciati}, {Tasca}, {Carollo}, {Silverman}, {Kampczyk}, {de Ravel},
  {Sanders}, {Scoville}, {Contini}, {Mainieri}, {Scodeggio}, {Kneib}, {Le
  F{\`e}vre}, {Bardelli}, {Bongiorno}, {Caputi}, {Coppa}, {de la Torre},
  {Franzetti}, {Garilli}, {Lamareille}, {Le Borgne}, {Le Brun}, {Mignoli},
  {Perez Montero}, {Pello}, {Ricciardelli}, {Tanaka}, {Tresse}, {Vergani},
  {Welikala}, {Zucca}, {Oesch}, {Abbas}, {Barnes}, {Bordoloi}, {Bottini},
  {Cappi}, {Cassata}, {Cimatti}, {Fumana}, {Hasinger}, {Koekemoer},
  {Leauthaud}, {Maccagni}, {Marinoni}, {McCracken}, {Memeo}, {Meneux}, {Nair},
  {Porciani}, {Presotto}, \& {Scaramella}}]{Peng10}
{Peng}, Y.-j., {Lilly}, S.~J., {Kova{\v c}}, K., {et~al.} 2010, \apj, 721, 193

\bibitem[{{Poggianti} {et~al.}(2016){Poggianti}, {Fasano}, {Omizzolo},
  {Gullieuszik}, {Bettoni}, {Moretti}, {Paccagnella}, {Jaff{\'e}}, {Vulcani},
  {Fritz}, {Couch}, \& {D'Onofrio}}]{Poggianti16}
{Poggianti}, B.~M., {Fasano}, G., {Omizzolo}, A., {et~al.} 2016, \aj, 151, 78

\bibitem[{{Poggianti} {et~al.}(2017){Poggianti}, {Moretti}, {Gullieuszik},
  {Fritz}, {Jaff{\'e}}, {Bettoni}, {Fasano}, {Bellhouse}, {Hau}, {Vulcani},
  {Biviano}, {Omizzolo}, {Paccagnella}, {D'Onofrio}, {Cava}, {Sheen}, {Couch},
  \& {Owers}}]{Poggianti17a}
{Poggianti}, B.~M., {Moretti}, A., {Gullieuszik}, M., {et~al.} 2017, \apj, 844,
  48

\bibitem[{{Ricciardelli} {et~al.}(2014){Ricciardelli}, {Cava}, {Varela}, \&
  {Quilis}}]{Ricciardelli14}
{Ricciardelli}, E., {Cava}, A., {Varela}, J., \& {Quilis}, V. 2014, \mnras,
  445, 4045

\bibitem[{{Rodighiero} {et~al.}(2011){Rodighiero}, {Daddi}, {Baronchelli},
  {Cimatti}, {Renzini}, {Aussel}, {Popesso}, {Lutz}, {Andreani}, {Berta},
  {Cava}, {Elbaz}, {Feltre}, {Fontana}, {F{\"o}rster Schreiber},
  {Franceschini}, {Genzel}, {Grazian}, {Gruppioni}, {Ilbert}, {Le Floch},
  {Magdis}, {Magliocchetti}, {Magnelli}, {Maiolino}, {McCracken}, {Nordon},
  {Poglitsch}, {Santini}, {Pozzi}, {Riguccini}, {Tacconi}, {Wuyts}, \&
  {Zamorani}}]{Rodighiero11}
{Rodighiero}, G., {Daddi}, E., {Baronchelli}, I., {et~al.} 2011, \apjl, 739,
  L40

\bibitem[{{Rudnick} {et~al.}(2017){Rudnick}, {Hodge}, {Walter}, {Momcheva},
  {Tran}, {Papovich}, {da Cunha}, {Decarli}, {Saintonge}, {Willmer}, {Lotz}, \&
  {Lentati}}]{Rudnick17}
{Rudnick}, G., {Hodge}, J., {Walter}, F., {et~al.} 2017, \apj, 849, 27

\bibitem[{{Saintonge} {et~al.}(2011{\natexlab{a}}){Saintonge}, {Kauffmann},
  {Wang}, {Kramer}, {Tacconi}, {Buchbender}, {Catinella}, {Graci{\'a}-Carpio},
  {Cortese}, {Fabello}, {Fu}, {Genzel}, {Giovanelli}, {Guo}, {Haynes},
  {Heckman}, {Krumholz}, {Lemonias}, {Li}, {Moran}, {Rodriguez-Fernandez},
  {Schiminovich}, {Schuster}, \& {Sievers}}]{Saintonge11a}
{Saintonge}, A., {Kauffmann}, G., {Wang}, J., {et~al.} 2011{\natexlab{a}},
  \mnras, 415, 61

\bibitem[{{Saintonge} {et~al.}(2011{\natexlab{b}}){Saintonge}, {Kauffmann},
  {Kramer}, {Tacconi}, {Buchbender}, {Catinella}, {Fabello},
  {Graci{\'a}-Carpio}, {Wang}, {Cortese}, {Fu}, {Genzel}, {Giovanelli}, {Guo},
  {Haynes}, {Heckman}, {Krumholz}, {Lemonias}, {Li}, {Moran},
  {Rodriguez-Fernandez}, {Schiminovich}, {Schuster}, \&
  {Sievers}}]{Saintonge11}
{Saintonge}, A., {Kauffmann}, G., {Kramer}, C., {et~al.} 2011{\natexlab{b}},
  \mnras, 415, 32

\bibitem[{{Salim} {et~al.}(2007){Salim}, {Rich}, {Charlot}, {Brinchmann},
  {Johnson}, {Schiminovich}, {Seibert}, {Mallery}, {Heckman}, {Forster},
  {Friedman}, {Martin}, {Morrissey}, {Neff}, {Small}, {Wyder}, {Bianchi},
  {Donas}, {Lee}, {Madore}, {Milliard}, {Szalay}, {Welsh}, \& {Yi}}]{Salim07}
{Salim}, S., {Rich}, R.~M., {Charlot}, S., {et~al.} 2007, \apjs, 173, 267

\bibitem[{{Santini} {et~al.}(2014){Santini}, {Maiolino}, {Magnelli}, {Lutz},
  {Lamastra}, {Li Causi}, {Eales}, {Andreani}, {Berta}, {Buat}, {Cooray},
  {Cresci}, {Daddi}, {Farrah}, {Fontana}, {Franceschini}, {Genzel}, {Granato},
  {Grazian}, {Le Floc'h}, {Magdis}, {Magliocchetti}, {Mannucci}, {Menci},
  {Nordon}, {Oliver}, {Popesso}, {Pozzi}, {Riguccini}, {Rodighiero}, {Rosario},
  {Salvato}, {Scott}, {Silva}, {Tacconi}, {Viero}, {Wang}, {Wuyts}, \&
  {Xu}}]{Santini14}
{Santini}, P., {Maiolino}, R., {Magnelli}, B., {et~al.} 2014, \aap, 562, A30

\bibitem[{{Sargent} {et~al.}(2013){Sargent}, {Daddi}, {B{\'e}thermin}, \&
  {Elbaz}}]{Sargent13}
{Sargent}, M.~T., {Daddi}, E., {B{\'e}thermin}, M., \& {Elbaz}, D. 2013,
  Asociacion Argentina de Astronomia La Plata Argentina Book Series, 4, 116

\bibitem[{{Schinnerer} {et~al.}(2016){Schinnerer}, {Groves}, {Sargent},
  {Karim}, {Oesch}, {Magnelli}, {LeFevre}, {Tasca}, {Civano}, {Cassata}, \&
  {Smol{\v c}i{\'c}}}]{Schinnerer16}
{Schinnerer}, E., {Groves}, B., {Sargent}, M.~T., {et~al.} 2016, \apj, 833, 112

\bibitem[{{Scott} {et~al.}(2013){Scott}, {Usero}, {Brinks}, {Boselli},
  {Cortese}, \& {Bravo-Alfaro}}]{Scott13}
{Scott}, T.~C., {Usero}, A., {Brinks}, E., {et~al.} 2013, \mnras, 429, 221

\bibitem[{{Scoville} {et~al.}(2007){Scoville}, {Aussel}, {Brusa}, {Capak},
  {Carollo}, {Elvis}, {Giavalisco}, {Guzzo}, {Hasinger}, {Impey}, {Kneib},
  {LeFevre}, {Lilly}, {Mobasher}, {Renzini}, {Rich}, {Sanders}, {Schinnerer},
  {Schminovich}, {Shopbell}, {Taniguchi}, \& {Tyson}}]{Scoville07}
{Scoville}, N., {Aussel}, H., {Brusa}, M., {et~al.} 2007, \apjs, 172, 1

\bibitem[{{Scoville} {et~al.}(2013){Scoville}, {Arnouts}, {Aussel}, {Benson},
  {Bongiorno}, {Bundy}, {Calvo}, {Capak}, {Carollo}, {Civano}, {Dunlop},
  {Elvis}, {Faisst}, {Finoguenov}, {Fu}, {Giavalisco}, {Guo}, {Ilbert},
  {Iovino}, {Kajisawa}, {Kartaltepe}, {Leauthaud}, {Le F{\`e}vre}, {LeFloch},
  {Lilly}, {Liu}, {Manohar}, {Massey}, {Masters}, {McCracken}, {Mobasher},
  {Peng}, {Renzini}, {Rhodes}, {Salvato}, {Sanders}, {Sarvestani}, {Scarlata},
  {Schinnerer}, {Sheth}, {Shopbell}, {Smol{\v c}i{\'c}}, {Taniguchi}, {Taylor},
  {White}, \& {Yan}}]{Scoville13}
{Scoville}, N., {Arnouts}, S., {Aussel}, H., {et~al.} 2013, \apjs, 206, 3

\bibitem[{{Scoville} {et~al.}(2014){Scoville}, {Aussel}, {Sheth}, {Scott},
  {Sanders}, {Ivison}, {Pope}, {Capak}, {Vanden Bout}, {Manohar}, {Kartaltepe},
  {Robertson}, \& {Lilly}}]{Scoville14}
{Scoville}, N., {Aussel}, H., {Sheth}, K., {et~al.} 2014, \apj, 783, 84

\bibitem[{{Scoville} {et~al.}(2017){Scoville}, {Lee}, {Vanden Bout},
  {Diaz-Santos}, {Sanders}, {Darvish}, {Bongiorno}, {Casey}, {Murchikova},
  {Koda}, {Capak}, {Vlahakis}, {Ilbert}, {Sheth}, {Morokuma-Matsui}, {Ivison},
  {Aussel}, {Laigle}, {McCracken}, {Armus}, {Pope}, {Toft}, \&
  {Masters}}]{Scoville17}
{Scoville}, N., {Lee}, N., {Vanden Bout}, P., {et~al.} 2017, \apj, 837, 150

\bibitem[{{Serra} {et~al.}(2012){Serra}, {Oosterloo}, {Morganti}, {Alatalo},
  {Blitz}, {Bois}, {Bournaud}, {Bureau}, {Cappellari}, {Crocker}, {Davies},
  {Davis}, {de Zeeuw}, {Duc}, {Emsellem}, {Khochfar}, {Krajnovi{\'c}},
  {Kuntschner}, {Lablanche}, {McDermid}, {Naab}, {Sarzi}, {Scott}, {Trager},
  {Weijmans}, \& {Young}}]{Serra12}
{Serra}, P., {Oosterloo}, T., {Morganti}, R., {et~al.} 2012, \mnras, 422, 1835

\bibitem[{{Sobral} {et~al.}(2011){Sobral}, {Best}, {Smail}, {Geach},
  {Cirasuolo}, {Garn}, \& {Dalton}}]{Sobral11}
{Sobral}, D., {Best}, P.~N., {Smail}, I., {et~al.} 2011, \mnras, 411, 675

\bibitem[{{Sobral} {et~al.}(2015){Sobral}, {Stroe}, {Dawson}, {Wittman}, {Jee},
  {R{\"o}ttgering}, {van Weeren}, \& {Br{\"u}ggen}}]{Sobral15}
{Sobral}, D., {Stroe}, A., {Dawson}, W.~A., {et~al.} 2015, \mnras, 450, 630

\bibitem[{{Sobral} {et~al.}(2016){Sobral}, {Stroe}, {Koyama}, {Darvish},
  {Calhau}, {Afonso}, {Kodama}, \& {Nakata}}]{Sobral16}
{Sobral}, D., {Stroe}, A., {Koyama}, Y., {et~al.} 2016, \mnras, 458, 3443

\bibitem[{{Speagle} {et~al.}(2014){Speagle}, {Steinhardt}, {Capak}, \&
  {Silverman}}]{Speagle14}
{Speagle}, J.~S., {Steinhardt}, C.~L., {Capak}, P.~L., \& {Silverman}, J.~D.
  2014, \apjs, 214, 15

\bibitem[{{Stark} {et~al.}(1986){Stark}, {Knapp}, {Bally}, {Wilson}, {Penzias},
  \& {Rowe}}]{Stark86}
{Stark}, A.~A., {Knapp}, G.~R., {Bally}, J., {et~al.} 1986, \apj, 310, 660

\bibitem[{{Tacconi} {et~al.}(2010){Tacconi}, {Genzel}, {Neri}, {Cox}, {Cooper},
  {Shapiro}, {Bolatto}, {Bouch{\'e}}, {Bournaud}, {Burkert}, {Combes},
  {Comerford}, {Davis}, {Schreiber}, {Garcia-Burillo}, {Gracia-Carpio}, {Lutz},
  {Naab}, {Omont}, {Shapley}, {Sternberg}, \& {Weiner}}]{Tacconi10}
{Tacconi}, L.~J., {Genzel}, R., {Neri}, R., {et~al.} 2010, \nat, 463, 781

\bibitem[{{Tacconi} {et~al.}(2013){Tacconi}, {Neri}, {Genzel}, {Combes},
  {Bolatto}, {Cooper}, {Wuyts}, {Bournaud}, {Burkert}, {Comerford}, {Cox},
  {Davis}, {F{\"o}rster Schreiber}, {Garc{\'{\i}}a-Burillo}, {Gracia-Carpio},
  {Lutz}, {Naab}, {Newman}, {Omont}, {Saintonge}, {Shapiro Griffin}, {Shapley},
  {Sternberg}, \& {Weiner}}]{Tacconi13}
{Tacconi}, L.~J., {Neri}, R., {Genzel}, R., {et~al.} 2013, \apj, 768, 74

\bibitem[{{Tacconi} {et~al.}(2018){Tacconi}, {Genzel}, {Saintonge}, {Combes},
  {Garc{\'{\i}}a-Burillo}, {Neri}, {Bolatto}, {Contini}, {F{\"o}rster
  Schreiber}, {Lilly}, {Lutz}, {Wuyts}, {Accurso}, {Boissier}, {Boone},
  {Bouch{\'e}}, {Bournaud}, {Burkert}, {Carollo}, {Cooper}, {Cox}, {Feruglio},
  {Freundlich}, {Herrera-Camus}, {Juneau}, {Lippa}, {Naab}, {Renzini},
  {Salome}, {Sternberg}, {Tadaki}, {{\"U}bler}, {Walter}, {Weiner}, \&
  {Weiss}}]{Tacconi18}
{Tacconi}, L.~J., {Genzel}, R., {Saintonge}, A., {et~al.} 2018, \apj, 853, 179

\bibitem[{{Tomczak} {et~al.}(2017){Tomczak}, {Lemaux}, {Lubin}, {Gal}, {Wu},
  {Holden}, {Kocevski}, {Mei}, {Pelliccia}, {Rumbaugh}, \& {Shen}}]{Tomczak17}
{Tomczak}, A.~R., {Lemaux}, B.~C., {Lubin}, L.~M., {et~al.} 2017, \mnras, 472,
  3512

\bibitem[{{Tran} {et~al.}(2015){Tran}, {Nanayakkara}, {Yuan}, {Kacprzak},
  {Glazebrook}, {Kewley}, {Momcheva}, {Papovich}, {Quadri}, {Rudnick},
  {Saintonge}, {Spitler}, {Straatman}, \& {Tomczak}}]{Tran15}
{Tran}, K.-V.~H., {Nanayakkara}, T., {Yuan}, T., {et~al.} 2015, \apj, 811, 28

\bibitem[{{Vogelsberger} {et~al.}(2014){Vogelsberger}, {Genel}, {Springel},
  {Torrey}, {Sijacki}, {Xu}, {Snyder}, {Bird}, {Nelson}, \&
  {Hernquist}}]{Vogelsberger14}
{Vogelsberger}, M., {Genel}, S., {Springel}, V., {et~al.} 2014, \nat, 509, 177

\bibitem[{{von der Linden} {et~al.}(2010){von der Linden}, {Wild}, {Kauffmann},
  {White}, \& {Weinmann}}]{Vonderlinden10}
{von der Linden}, A., {Wild}, V., {Kauffmann}, G., {White}, S.~D.~M., \&
  {Weinmann}, S. 2010, \mnras, 404, 1231

\bibitem[{{Vulcani} {et~al.}(2010){Vulcani}, {Poggianti}, {Finn}, {Rudnick},
  {Desai}, \& {Bamford}}]{Vulcani10}
{Vulcani}, B., {Poggianti}, B.~M., {Finn}, R.~A., {et~al.} 2010, \apjl, 710, L1

\bibitem[{{Wagg} {et~al.}(2012){Wagg}, {Pope}, {Alberts}, {Armus}, {Brodwin},
  {Bussmann}, {Desai}, {Dey}, {Jannuzi}, {Le Floc'h}, {Melbourne}, \&
  {Stern}}]{Wagg12}
{Wagg}, J., {Pope}, A., {Alberts}, S., {et~al.} 2012, \apj, 752, 91

\bibitem[{{Wang} {et~al.}(2018){Wang}, {Norberg}, {Brough}, {Brown}, {da
  Cunha}, {Davies}, {Driver}, {Holwerda}, {Hopkins}, {Lara-Lopez}, {Liske},
  {Loveday}, {Grootes}, {Popescu}, \& {Wright}}]{Wang18}
{Wang}, L., {Norberg}, P., {Brough}, S., {et~al.} 2018, \aap, 618, A1

\bibitem[{{Wijesinghe} {et~al.}(2012){Wijesinghe}, {Hopkins}, {Brough},
  {Taylor}, {Norberg}, {Bauer}, {Brown}, {Cameron}, {Conselice}, {Croom},
  {Driver}, {Grootes}, {Jones}, {Kelvin}, {Loveday}, {Pimbblet}, {Popescu},
  {Prescott}, {Sharp}, {Baldry}, {Sadler}, {Liske}, {Robotham}, {Bamford},
  {Bland-Hawthorn}, {Gunawardhana}, {Meyer}, {Parkinson}, {Drinkwater},
  {Peacock}, \& {Tuffs}}]{Wijesinghe12}
{Wijesinghe}, D.~B., {Hopkins}, A.~M., {Brough}, S., {et~al.} 2012, \mnras,
  423, 3679

\bibitem[{{Young} \& {Scoville}(1991)}]{Young91}
{Young}, J.~S., \& {Scoville}, N.~Z. 1991, \araa, 29, 581

\end{thebibliography}

\end{document}